\documentclass[aps,pre,showpacs,superscriptaddress,twocolumn]{revtex4-1}
\usepackage{graphicx}
\usepackage{amsmath}
\usepackage{amssymb}
\usepackage{mathtools}
\usepackage{makeidx}
\usepackage{amsfonts}
\usepackage{bm}
\usepackage[pdftex]{hyperref}
\hypersetup{colorlinks=true,linkcolor=blue,citecolor=blue,urlcolor=red}

\begin{document}

\title{Symmetry and its breaking in path integral approach to quantum Brownian motion}

\author{Joonhyun Yeo}\email{jhyeo@konkuk.ac.kr}
\affiliation{Department of Physics,
Konkuk University, Seoul 05029, Korea}
\date{\today}

\begin{abstract}
We study the Caldeira-Leggett model where a quantum Brownian particle 
interacts with an environment or a bath consisting
of a collection of harmonic oscillators in the path integral formalism. 
Compared to the contours that the paths take in the conventional Schwinger-Keldysh
formalism, the paths in our study are deformed in the complex time plane as suggested by the 
recent study [C. Aron, G. Biroli and L. F. Cugliandolo, SciPost Phys.\ {\bf 4}, 008 (2018)].
This is done to investigate the connection between 
the symmetry properties in the Schwinger-Keldysh action 
and the equilibrium or non-equilibrium nature of the dynamics in an open quantum system.
We derive the influence functional explicitly in this setting, which captures the effect of the coupling to the bath.
We show that in equilibrium the action and the influence functional are invariant under a set of transformations of
path integral variables. The fluctuation-dissipation relation is obtained as a consequence of this symmetry. When
the system is driven by an external time-dependent protocol, the symmetry is broken. From the terms that break
the symmetry, we derive a quantum Jarzynski-like equality for a quantum mechanical work-like quantity 
given as a function of 
fluctuating quantum trajectory. In the classical limit, the transformations becomes those used in the functional integral
formalism of the classical stochastic thermodynamics to derive the classical fluctuation theorem.  
\end{abstract}



\maketitle
\section{Introduction}
\label{sec:Introduction}

Understanding the behavior of thermodynamic quantities and their fluctuations
in non-equilibrium situations is a major challenge in statistical mechanics.
Recent development of fluctuation theorems (FTs)~\cite{evans,gallavotti,jarzynski,crooks,kurchan,lebowitz}
has provided crucial insights into this problem. 
In classical systems, theoretical progress has been made via
stochastic thermodynamics \cite{sekimoto,seifert_review}, where thermodynamic quantities, such as
work, heat or entropy production, are
attributed to an individual stochastic trajectory. 
According to this approach, a probability can be assigned to each stochastic path,
and by investigating how it changes under the time-reversed dynamics, one can identify
many different forms of FTs arising in
various physical situations 
\cite{crooks,jarzynski,kurchan,hatano,speck,seifert,esposito,chernyak,spinney,lkp,gen_adj,kylp,yklp}. 
Attempts to extend the classical FTs to the quantum regime have been made \cite{kurchan1,tasaki} resulting in
the quantum version of FTs \cite{esposito1,campisi,ht}. A most notable example is 
the quantum FT for the fluctuating work defined in the two projective 
energy measurement scheme (TPM) \cite{talkner1,campisi1,talkner2}. 
However, there exist many other definitions of quantum work 
\cite {chernyak1,subasi,deffner,campisi2,alla,deffner2,venka,solinas1,alonso,kammer,talkner3,deffner3,miller,solinas,hofer,sampaio} 
and an appropriate definition for the quantum work is still under debate \cite{nogo,baumer}. 
This difficulty stems partly from the fact that,
unlike classical stochastic thermodynamics, there is no clear notion of a trajectory in quantum systems.
 
In classical stochastic thermodynamics, the path probability for a stochastic path is given by the 
Onsager-Machlup (OM) form \cite{onsager}. By considering the so-called irreversibility 
defined by the logarithm of the ratio of the probabilities for the forward and the time-reversed paths, one can 
derive the various FTs mentioned above. An alternative approach is to study the
symmetry properties of the Martin-Siggia-Rose-Janssen-De Dominicis (MSRJD) 
functional \cite{msr,janssen,dedom} for the stochastic dynamics,
where an auxiliary response variable is introduced in addition to the dynamical variable. 
Although the MSRJD formalism is
equivalent to the OM one in the sense that, when integrated
over this response variable, one recovers the OM form, it provides 
an additional useful information on equilibrium and non-equilibrium dynamics.
In this formalism, one considers
a set of time reversal transformations for the main and the auxiliary variables. 
Equilibrium is characterized by the invariance of the MSRJD functional 
under this transformation \cite{lubensky,andreanov,mallick,abc1,abc2,arenas,kim}, and
the fluctuation-dissipation relation (FDR) follows as a Ward-Takahashi 
identity of this symmetry \cite{andreanov,mallick,abc1}. 
In non-equilibrium situations, the symmetry is broken. The term responsible for the 
breaking of the symmetry can be used to derive various FTs \cite{mallick,abc1}. 

Given the success of the classical stochastic thermodynamics, it is natural to seek out a quantum version
of stochastic thermodynamics. There have been 
many recent attempts \cite{aurell,carrega1,carrega,funo,funo1,qiu} 
to extend it to the quantum regime using the path integral method \cite{feynman}, which
can be regarded as the natural quantum generalization of those using a classical stochastic path. 
We note that all these approaches \cite{aurell,carrega1,carrega,funo,funo1,qiu}
are confined to treating thermodynamic quantities defined in the TPM scheme.   
Another route, which we take in this paper, is to look for the quantum generalization
of the classical method that uses the
symmetry properties of the MSRJD functional integral formalism \cite{lubensky,andreanov,mallick,abc1,abc2,arenas,kim}
within the path integral 
formalism .
A work in this direction was carried out in Ref.~\cite{sieb},
where the quantum version of the field transformations was identified
in the Schwinger-Keldysh (SK) path integral formalism \cite{kamenev,stef}. The SK action in equilibrium is 
shown to be invariant under this transformation, and the FDR is obtained
as a corollary of this symmetry \cite{sieb}. This was generalized in Ref.~\cite{abc} to 
non-equilibrium dynamics of a {\it closed} quantum system. 
In order to find the field transformation for the dynamics in 
a finite time interval, it is necessary to formulate the SK path integral 
on a deformed contour on the complex time plane \cite{abc}
compared to the standard Kadanoff-Baym one \cite{stef}.  
From the symmetry breaking term out of equilibrium, the quantum FT for the quantum work, 
which is not based on the TPM scheme, is obtained \cite{abc}. 

The purpose of this paper is to explore further the latter approach to quantum thermodynamics 
which uses the symmetry properties of the SK path integral formalism, and
extend it to an {\it open} quantum system, where the system interacts with an environment. 
As a paradigmatic model of an open quantum system,
we study the path integral formulation of the quantum Brownian motion \cite{grabert}
using the Caldeira-Leggett model \cite{cl}, where the environment is represented
by a collection of harmonic oscillators. 
By applying the method developed in Ref.~\cite{abc},
we investigate how the existence of an environment 
is encoded in the path integral formulation especially in the symmetry properties of the SK action. 
On the deformed time contour, we identify the SK action
due to the presence of the environment, known as the influence functional \cite{fv}, 
responsible for the dissipation into the environment. We find the field transformations that
leave the action invariant in equilibrium.  We show explicitly that 
the transformations reduce in the classical limit
to those used in classical stochastic thermodynamics thereby making an explicit connection with
the MSRJD formalism of classical stochastic thermodynamics. 
When the system is driven out of equilibrium, the symmetry is broken from which
we establish quantum FT for a quantum mechanical work-like quantity which depend on the quantum SK trajectory
as in Ref.~\cite{abc}, which is a generalization of the classical work defined in stochastic thermodynamics.

In the next section, we present the SK path integral formalism for the Caldeira-Leggett model on the deformed time contour
following the procedure in Ref.~\cite{abc}. In Sec.~\ref{sec:eq}, we identify the field transformations that leave the action invariant.
We also show that the equilibrium FDR follows from this symmetry. In Sec.~\ref{sec:noneq},
we consider the case where the system is driven out of equilibrium. By identifying the term that breaks the symmetry,
we derive the quantum Jarzynski equality. We then apply this analysis to a concrete example
where the system is in a harmonic potential whose center is pulled in a time-dependent manner.
In the following section, we show explicitly that the present formalism 
reduces to the MSRJD one for the generalized Langevin equation. We then summarize and conclude with discussion.

\section{Path Integral on Deformed Contours}
\label{sec:path}

In this section, we briefly review the basic idea behind the contour deformation proposed in Ref.~\cite{abc},
and apply it to the path integral representation of the quantum Brownian motion. 
We consider the Caldeira-Leggett model \cite{cl}, where 
a quantum mechanical system interacts with an environment or a bath, which consists of
a collection of harmonic oscillators. The total Hamiltonian is given by 
$H_{\rm tot}(t)=H_{\rm S}+H_{\rm B}+H_{\rm I}$.
The explicit time dependence of the Hamiltonian comes from 
the system Hamiltonian given by
  \begin{equation}
   H_{\rm S}=\frac{p^2}{2m}+V(x,\lambda_t )
  \end{equation}
with the potential energy depending on an external time-dependent protocol $\lambda_t$.
The bath is a collection of harmonic oscillators with frequencies $\omega_n$ ($n=1,2,\cdots$):
\begin{equation}
 H_{\rm B} = \sum_n \left( \frac{p_n^2}{2m_n}+\frac{1}{2}m_n\omega^2_n q^2_n \right),
\end{equation}
and the system and the bath interact via the interaction Hamiltonian
\begin{equation} 
H_{\rm I}=-x \sum_n c_n q_n +\frac{\mu}{2} x^2,
\end{equation} 
 where 
  \begin{equation}
\mu\equiv\sum_n\frac{c^2_n}{m_n\omega^2_n},
\end{equation}
and the last term incorporates the renormalization of the system Hamiltonian due to the coupling to the bath \cite{cl}.

The time evolution of density operator of the total system is given by 
$\rho(t)=U_{t,0}\rho(0)U^\dagger_{t,0}$ where
\begin{equation}
U_{t,0}=\mathbb{T}\exp(-\frac{i}{\hbar}\int_0^t H_{\rm tot}(s)ds)
\end{equation}
with the time ordering operator $\mathbb{T}$. 
In the following, we develop the path integral formalism for 
the density matrix and correlation functions.  
The time evolution of the system will then be described
by tracing out the bath degrees of freedom. 
For the path integral formalism, we first write 
the completeness relation for the states $\vert\bm{Q}\rangle\equiv\vert x,\bm{q}\rangle\equiv\vert x,q_1,q_2,\cdots\rangle $,
\begin{equation}
 1=\int d{\bm Q}\; | \bm{Q}\rangle\langle \bm{Q} | .
\label{comp_0}
 \end{equation}
The density operator can be rewritten as
\begin{align}
\rho(t) =&  \int d\bm{Q}_f 
d\bm{Q}^\prime_f d\bm{Q}_i d\bm{Q}^\prime_i\;  |\bm{Q}_f\rangle\langle \bm{Q}_f  |U_{t,0}|\bm{Q}_i\rangle \nonumber \\
&\times\langle \bm{Q}_i | \rho(0)|\bm{Q}^\prime_i \rangle 
\langle \bm{Q}^\prime_i | U^\dag_{t,0} |\bm{Q}'_f\rangle
\langle \bm{Q}'_f | , \label{rho_r}
\end{align} 
where $\bm{Q}_f$ stands for $(x_f,\bm{q}_f)$, etc.
The standard path integral representation \cite{feynman,grabert, kleinert}
is obtained by inserting the completeness relation, Eq.~(\ref{comp_0}) at time slices  
appearing in the discretized expressions of $U$ and $U^\dag$.
This can be rewritten in terms of the path integral over 
the paths $x_{\pm}(s)$ and $\bm{q}_{\pm}(s)$, $0\le s\le t$, of the system 
and bath particles, respectively. The 
forward paths $x_+$ and $\bm{q}_+$ arising from the matrix element of $U$
have end points $x_+(0)=x_i$, $x_+(t)=x_f$, $\bm{q}_+(0)=\bm{q}_i$
and $\bm{q}_+(t)=\bm{q}_f$. On the other hand, 
the backward paths corresponding to $U^\dagger$ have end points
$x_-(0)=x'_i$, $x_-(t)=x'_f$, $\bm{q}_-(0)=\bm{q}_i$
and $\bm{q}_-(t)=\bm{q}'_f$. 
In this paper, we consider the
case where the system and the bath are initially 
at equilibrium, i.e.
\begin{equation}
\rho(0)=\frac 1 {Z_\beta(0)} e^{-\beta H_{\rm tot}(0)},
\label{init}
\end{equation}
where $Z_\beta(0)=\mathrm{Tr}e^{-\beta H_{\rm tot}(0)}$.
We note that the matrix element involving $\rho(0)$ can be represented by the
path integral for paths running along the imaginary time axis.
In this way, the standard path integral is represented on the Kadanoff-Baym contour \cite{stef}
consisting of the forward, backward and imaginary-time branches.

As mentioned in Introduction, we use the path 
integral representation on a deformed time contour \cite{abc} instead of the standard one.
The key element of this formulation is the use of an alternative completeness relation,
instead of Eq.~(\ref{comp_0}), given by 
\begin{equation}
 1=\int d{\bm Q}\; e^{\frac i\hbar \theta(s)H_{\rm tot}(s)}
 |\bm{Q}\rangle  \langle \bm{Q} | e^{-\frac i\hbar \theta(s)H_{\rm tot}(s)},  \label{comp}
\end{equation}
at an arbitrary time $s$, where we have used phase factors characterized by
an arbitrary complex-valued function $\theta(s)$ of time $s$. 
As we will see below, the actual form of $\theta(s)$
determines the deformation of the paths
$x(z)$, $\bm{q}(z)$ on the complex time $z$-plane.
In this paper, as in Ref.~\cite{abc}, two different functions
$\theta_+(s)$ and $\theta_-(s)$ for forward and backward paths, respectively will be used.
Using these functions and the completeness relation Eq.~(\ref{comp}) in Eq.~(\ref{rho_r}), 
we can write the normalization
of the density operator $1=\mathrm{Tr}\rho(t)$ as  
\begin{widetext}
\begin{align}
 1  = &\int d\bm{Q}_f d\bm{Q}'_f
 d\bm{Q}_i d\bm{Q}'_f   \;
  \langle \bm{Q}'_f |e^{-\frac i \hbar \theta_-(t)H_{\rm tot}(t)}  
 e^{\frac i \hbar \theta_+(t) H_{\rm tot}(t)}
 |\bm{Q}_f\rangle  \langle \bm{Q}_f |e^{-\frac i \hbar \theta_+(t)H_{\rm tot}(t)}\;U_{t,0}\; e^{\frac i\hbar \theta_+(0)H_{\rm tot}(0)}
 |\bm{Q}_i\rangle \nonumber \\ 
 &\quad\quad\quad\quad\quad 
 \times \langle \bm{Q}_i |e^{-\frac i\hbar \theta_+(0)H_{\rm tot}(0)}\; \rho(0)\;e^{\frac i\hbar \theta_-(0)H_{\rm tot}(0)}
 |\bm{Q}'_i\rangle  \langle \bm{Q}'_i |e^{-\frac i\hbar \theta_-(0)H_{\rm tot}(0)}\; U^\dag_{t,0}\; e^{\frac i\hbar \theta_-(t)H_{\rm tot}(t)}
 |\bm{Q}'_f\rangle . \label{trace}
\end{align}
\end{widetext}
We are also interested in expressing the expectation values of system operators 
in terms of the path integrals over the deformed 
time contours. For example, for a {\it system} operator $A_{\rm S}$, we have for $0\le t_1\le t$,
$\langle A_{\rm S} (t_1)\rangle=\mathrm{Tr}(A_{\rm S}\rho(t_1))=\mathrm{Tr}(A_{\rm S}(t_1)\rho(0))$,
where $A_{\rm S}(t_1)=U^\dag_{t_1,0}A_{\rm S} U_{t_1,0}$ is the Heisenberg operator. 
This can also be rewritten as $\langle A_{\rm S} (t_1)\rangle=\mathrm{Tr}(U^\dag_{t,0}U_{t,t_1}
A_{\rm S} U_{t_1,0}\rho(0))$, and if we use the completeness relation, Eq.~(\ref{comp}), we have
\begin{widetext}
\begin{align}
 \langle A_{\rm S} (t_1) \rangle &= \int d\bm{Q}_f d\bm{Q}'_f
 d\bm{Q}_i d\bm{Q}'_f d\bm{Q}_1 d\bm{Q}'_1   
 \langle \bm{Q}'_f |e^{-\frac i \hbar \theta_-(t)H_{\rm tot}(t)}  
 e^{\frac i \hbar \theta_+(t) H_{\rm tot}(t)}
 |\bm{Q}_f\rangle  
 \langle \bm{Q}_f |e^{-\frac i \hbar \theta_+(t)H_{\rm tot}(t)}\;U_{t,t_1}\; e^{\frac i\hbar \theta_+(t_1)H_{\rm tot}(t_1)}
 |\bm{Q}'_1\rangle \nonumber \\ 
 &\quad\quad\quad\quad
 \times  \langle \bm{Q}'_1 |e^{-\frac i \hbar \theta_+(t_1)H_{\rm tot}(t_1)}\;A_{\rm S}\; e^{\frac i\hbar \theta_+(t_1)H_{\rm tot}(t_1)}
 |\bm{Q}_1\rangle 
  \langle \bm{Q}_1 |e^{-\frac i \hbar \theta_+(t_1)H_{\rm tot}(t_1)}\;U_{t_1,0}\; e^{\frac i\hbar \theta_+(0)H_{\rm tot}(0)}
 |\bm{Q}_i\rangle \nonumber \\
 &\quad\quad\quad\quad
 \times \langle \bm{Q}_i |e^{-\frac i\hbar \theta_+(0)H_{\rm tot}(0)}\; \rho(0)\;e^{\frac i\hbar \theta_-(0)H_{\rm tot}(0)}
 |\bm{Q}'_i\rangle 
 \langle \bm{Q}'_i |e^{-\frac i\hbar \theta_-(0)H_{\rm tot}(0)}\; U^\dag_{t,0}\; e^{\frac i\hbar \theta_-(t)H_{\rm tot}(t)}
 |\bm{Q}'_f\rangle . \label{one_point}
\end{align}
\end{widetext}

In the next section, we will also consider two-time correlation functions such as 
\begin{equation}
 \langle A_{\rm S}(t_1)
B_{\rm S}(t_2)\rangle=\mathrm{Tr}(U^\dag_{t_1,0}A_{\mathrm{S}} 
U^\dag_{t,t_1}U_{t,t_2}B_{\mathrm{S}}
U_{t_2,0}\rho(0))
\label{twotime}
\end{equation}
for two system operators $A_{\mathrm{S}}$ and $B_{\mathrm{S}}$. 
We can represent this quantity on a deformed contour as well
by using a similar expression to Eq.~(\ref{one_point}) in which $B_{\mathrm{S}}$ is inserted 
on the forward path in the form 
\begin{equation}
e^{- \frac i\hbar  \theta_+(t_2)H_{\rm tot}(t_2)}B_{\rm S} e^{ \frac i\hbar  \theta_+(t_2)H_{\rm tot}(t_2)}, 
\end{equation}
whereas $A_{\mathrm{S}}$ is 
on the backward path in the form 
\begin{equation}
e^{-\frac i\hbar \theta_-(t_1)H_{\rm tot}(t_1)}A_{\rm S} e^{\frac i\hbar \theta_-(t_1)H_{\rm tot}(t_1)}.
\end{equation}

In the following sections, we will develop path integral representation for these quantities
and study symmetries and broken symmetries for them.
In order to do that we discretize the time intervals appearing in $U$ and $U^\dag$
and insert the completeness relation, Eq.~(\ref{comp}) into each time slice. We then have to evaluate the matrix
element between the neighboring time steps to get the Lagrangians. After obtaining the action
for the total system, we will integrate over the bath variables to express everything in terms of the
system variables only.
As we will see below, 
the evaluation will take a quite different route depending on whether the Hamiltonian
has an explicit time dependence or not. We will discuss these two cases in detail below as well as the other parts. 

\section{Equilibrium}
\label{sec:eq}

In this section, we first develop the path integral formulation for 
the case where the Hamiltonian is time independent, i.e. $\partial_{\lambda_t} V=0$. 
As explained above, for the matrix element involving $U_{t,0}$ in Eq.~(\ref{trace}), 
we have to evaluate the matrix element between the neighboring discretized time steps $s_k$ and $s_{k+1}$,
\begin{align}
& \langle x_{k+1},\bm{q}_{k+1}|e^{-i\theta_+(s_{k+1})H_{\rm tot}/\hbar}e^{-iH_{\rm tot} ds/\hbar} \nonumber \\
&\quad\quad\quad\quad\quad \times e^{i\theta_+(s_{k})H_{\rm tot}/\hbar}|x_k,\bm{q}_k\rangle \nonumber \\
=& \langle x_{k+1},\bm{q}_{k+1}|e^{-i(1+\dot{\theta}_+(s_k))ds H_{\rm tot}/\hbar} |x_k,\bm{q}_k\rangle,
\label{mat_elem}
\end{align}
where $ds=s_{k+1}-s_k$. This form suggests a reparametrization of time into
a complex one $z_+(s)=s+\theta_+(s)$
so that $dz_+=(1+\dot{\theta}_+(s))ds$. 
The above matrix element can then be written as $\exp[(i/\hbar)dz_+ \mathcal{L}_+]$
in terms of the Lagrangian $\mathcal{L}_+$ given 
as a function of the paths $x_+(z_+), \bm{q}_+(z_+)$ and their velocities 
$\dot{x}_+=dx_+/dz_+$, $\dot{\bm{q}}_+=d\bm{q}_+/dz_+$ 
defined along the complex time $z_+$
with $x_k=x_+(z_+(s_k))$, $x_{k+1}=x_+(z_+(s_{k+1}))$, $\bm{q}_k=\bm{q}_+(z_+(s_k))$
and $\bm{q}_{k+1}=\bm{q}_+(z_+(s_{k+1}))$.
Collecting all these parts from the time slices, we end up with a path integral over the fluctuating paths,
$x_+(z_+), \bm{q}_+(z_+)$ of a factor 
$\exp[(i/\hbar)\int dz_+ \mathcal{L}_+ ]$, where the integral on the complex time plane is 
along the contour $z_+(s)$, $0\le s\le t$ in the direction
from $z_+(0)$ to $z_+(t)$ with the endpoints $x_+(z_+(0))=x_i$, $x_+(z_+(t))=x_f$,
$\bm{q}_+(z_+(0))=\bm{q}_i$, and $\bm{q}_+(z_+(t))=\bm{q}_f$. 

Similarly, the matrix element involving $U^\dag$ in Eq.~(\ref{trace}) is given by the path integral over the path 
$x_-(z_-), \bm{q}_-(z_-)$ defined along the complex time $z_-(s)=s+\theta_-(s)$ of a factor
$\exp[(i/\hbar)\int dz_- \mathcal{L}_- ]$. The integral, in this case, is along the contour
$z_-(s)$, $0\le s\le t$, starting from $z_-(t)$ ending at $z_-(0)$ (a backward path). 

The Lagrangians $\mathcal{L}_{\rm tot}^\pm$ have
three parts originating from the corresponding Hamiltonians,
$\mathcal{L}_{\rm tot}^\pm=\mathcal{L}^\pm_{\rm S}+\mathcal{L}^\pm_{\rm B}+\mathcal{L}^\pm_{\rm I}$, which are given respectively by
 \begin{align}
& \mathcal{L}^\pm_{\rm S}=\frac m 2 \left(\frac{dx_\pm}{dz_\pm}\right)^2-V(x_\pm(z_\pm))  , \label{LS}\\
& \mathcal{L}^\pm_{\rm B}= \sum_n \left[ 
\frac {m_n} 2 \left(\frac{dq_{\pm,n}}{dz_\pm}\right)^2-\frac 1 2 m_n \omega^2_n q^2_{\pm,n}(z_\pm)\right], \label{LB}\\
& \mathcal{L}^\pm_{\rm I}=x_\pm(z_\pm)\sum_n 
 c_n q_{\pm,n}(z_\pm) - \frac{\mu }{2} x^2_\pm(z_\pm). \label{LI}
 \end{align}

\begin{figure}
  \includegraphics[width=0.4\textwidth]{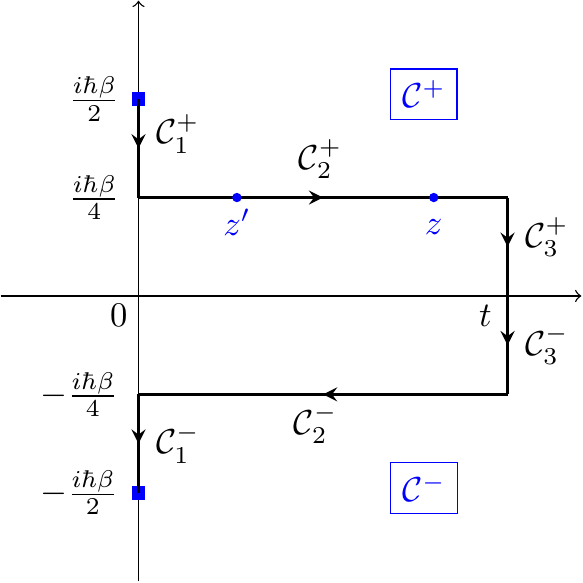}  
\caption{Contours $\mathcal{C}^\pm$ for 
$x_\pm(z)$ and $\bm{q}_{\pm}(z)$.  $\mathcal{C}^+$ runs from $i\hbar\beta/2$ to $t$
and $\mathcal{C}^-$ from $t$ to $-i\hbar\beta/2$. Each has three segments
denoted by $\mathcal{C}^\pm_1$,
$\mathcal{C}^\pm_2$ and $\mathcal{C}^\pm_3$. 
On the endpoints (squares), the paths
have the same value, over which the trace is performed.
For the calculation of the influence
functional $\Psi$ in Eq.~(\ref{psi}), 
the double contour integral over $z$ and $z'$ along $\mathcal{C}^{\pm}$ is 
performed such that $z$ is always ahead of $z'$.  }
\label{fig_one_point}
\end{figure}

So far the actual contours on the complex time plane are completely general and depends on the detailed form
of $\theta_{\pm}(s)$. 
In this paper, as in Ref.~\cite{abc}, we take a symmetric constant form, where 
\begin{equation}
\theta_{\pm}(s)=\pm \frac {i \hbar\beta}4
\end{equation}
for $0\le s\le t$. Other choices are possible \cite{abc}, but 
the present one is most convenient for the discussion on the symmetry properties 
of the actions. The contours are then just horizontal lines parallel to the real time axis given by 
$s\pm  i\hbar\beta/4$, $0\le s\le t$ on the complex-time plane 
as shown in Fig.~\ref{fig_one_point}. 
With this choice, the first factor in Eq.~(\ref{trace}) is just the matrix element of
$e^{-\beta H_{\rm tot}/2}$ which can be written as a path
integral over the paths which are given along the imaginary time axis (with the real part being equal to $t$). 
Finally, the third factor in Eq.~(\ref{trace}), which is the matrix element involving $\rho(0)$, 
again gives that of $e^{-\beta H_{\rm tot}/2}$.
We split this into $e^{-\beta H_{\rm tot}/4}e^{-\beta H_{\rm tot}/4}$. 
Then the paths in this case are given on the two parts along the imaginary 
time axis as shown Fig.~\ref{fig_one_point}, whose real parts are 0.

Combining all these, we find that we have to use the paths $x_\pm(z)$ and $\bm{q}_{\pm}(z)$ given along
the contours $\mathcal{C}^{\pm}$ on the complex time plane as shown  
Fig.~\ref{fig_one_point}. This is quite different from the standard Kadanoff-Baym contour \cite{grabert, stef,weiss}, but
is an entirely equivalent representation. 
The upper and lower branches, $\mathcal{C}^+$ and $\mathcal{C}^-$ 
run from $i\hbar\beta/2$ to $t$ and $t$ to $-i\hbar\beta/2$, respectively.
The normalization condition, Eq. (\ref{trace}) can then be written as 
\begin{align}
&1=\frac 1 {Z_\beta (0)}
\int dx_i \int d\bm{q}_i \int dx_f \int d\bm{q}_f \nonumber \\
&\times \int_{x_i}^{x_f}\mathcal{D}x_+(z)
\int_{x_f}^{x_i}
\mathcal{D}x_-(z)\int_{\bm{q}_i}^{\bm{q}_f}\mathcal{D}\bm{q}_+(z)\int_{\bm{q}_f}^{\bm{q}_i}\mathcal{D}\bm{q}_-(z)  \nonumber \\
&\times \exp\Big[ \frac i \hbar \sum_{a=+,-} \int_{\mathcal{C}^a} dz \; \mathcal{L}^a_{\rm tot}(x_a(z),\bm{q}_a(z))  
 \Big] , 
\label{normal_pi}
\end{align}
where the end points of the path integrals indicate 
the conditions that the paths are subject to, that is,
$x_+(i\hbar\beta/2)=x_-(-i\hbar\beta/2)=x_i$, $x_+(t)=x_-(t)=x_f$, $\bm{q}_+( i\hbar\beta/2)=\bm{q}_-( -i\hbar\beta/2)=\bm{q}_i$,
and $\bm{q}_+(t)=\bm{q}_-(t)=\bm{q}_f$.

We now integrate over the bath degrees of freedom
to express everything in terms of the system variables $x_\pm(z)$ only. 
The bath variables can explicitly integrated away, since the integrals are Gaussian.
The effect of the coupling to the bath then appears
as the influence functional \cite{fv, cl, grabert}. Upon integrating over $\bm{q}_{\pm}$ in Eq.~(\ref{normal_pi}), we obtain
\begin{align}
1=&\frac 1 {Z(0)}\int dx_i \int dx_f
\int_{x_i}^{x_f}\mathcal{D}x_+(z)
\int_{x_f}^{x_i}\mathcal{D}x_-(z) \nonumber \\
&\times  \exp\left(\frac{i}{\hbar}S_+[x_+]+\frac{i}{\hbar}S_-[x_-]-\frac{1}{\hbar}\Psi[x_+,x_-]\right),
 \label{S_ave1}
\end{align}
where 
\begin{equation}
S_{\pm}[x_{\pm}]=\int_{\mathcal{C}_{\pm}}dz\; \mathcal{L}^{\pm}_{\mathrm S}(x_\pm(z))
\label{action}
\end{equation} 
is the system action,
and $Z(0)=Z_\beta(0)/Z_{\rm B}$ with 
\begin{equation}
Z_{\rm B}=\mathrm{Tr}_{\rm B} e^{-\beta H_{\rm B}}=\prod_n \frac{1}{2\sinh(\beta\hbar\omega_n/2)}.
\label{zb}
\end{equation}
The effect of the coupling to the bath is reflected in Eq.~(\ref{S_ave1}) 
in the form of the influence functional $\Psi$, which we find after a lengthy algebra 
\begin{align}
\Psi[x_+,x_-] =&\int\limits_{\mathcal{C}^+} dz\int\limits_{\mathcal{C}^+,~z>z'} dz' \; x_+(z) K(z-z') x_+(z') \nonumber \\
+ &\int\limits_{\mathcal{C}^-} dz\int\limits_{\mathcal{C}^-,~z>z'} dz' \; x_-(z) K(z-z') x_-(z')  \nonumber\\
+&\int\limits_{\mathcal{C}^-} dz\int\limits_{\mathcal{C}^+} dz' \; x_-(z) K(z-z') x_+(z') \nonumber \\
+&i\frac{\mu}{2}\int\limits_{\mathcal{C}^+} dz\; x^2_+(z)+i\frac{\mu}{2}\int\limits_{\mathcal{C}^-} dz\; x^2_-(z), \label{psi}
 \end{align}
where
\begin{equation}    
K(z)\equiv\sum_n\frac{c^2_n}{2m_n\omega_n}
 \frac{\cosh(\frac 1 2\beta\hbar\omega_n -i\omega_n z)}{\sinh(\frac 1 2\beta\hbar\omega_n)}.
 \label{kernel}
  \end{equation}
In Eq.~(\ref{psi}), $z>z'$ indicates that the double contour integral is 
to be performed under the condition that $z'$ is behind $z$ as shown in Fig.~\ref{fig_one_point}.
The calculation of the influence functional involves evaluating the Gaussian path integrals and 
applying to the branches shown in Fig.~\ref{fig_one_point}. The details of this calculation is presented in 
Appendix \ref{sec:appA}.

\subsection{Equilibrium Symmetry}

We consider the change of variables in the path integral in Eq.~(\ref{S_ave1}) as follows:
\begin{equation}
x_{\pm}(z)\rightarrow \tilde{x}_{\pm}(z)\equiv x_{\pm}( t-z\pm \frac{i\hbar\beta}{2}) .
\label{trans}
\end{equation}
We investigate how the action $S_{\pm}$ and the influence functional $\Psi$ change under this
transformation. In order to do that, we first decompose the contour $\mathcal{C}^+$ into three parts, 
$\mathcal{C}_{1}^+,\mathcal{C}_{2}^+$ and $\mathcal{C}_{3}^+$ as shown in 
Fig.~\ref{fig_one_point}, where $z$ runs from $i\hbar\beta/2$ to $i\hbar\beta/4$,
from $i\hbar\beta/4$ to $t+i\hbar\beta/4$, and from $t+i\hbar\beta/4$ to $t$, respectively. 
The corresponding contributions
from these contours to the action $S_+$ in Eq.~(\ref{action}) 
are denoted by $S_{1}^+$, 
$S_{2}^+$ and $S_{3}^+$, respectively. 
Then, since $z=is$ with $s$ running from $\hbar\beta/2$ to $\hbar\beta/4$ on 
$\mathcal{C}_{1}^+$, we can write
\begin{align}
 S_{1}^+[\tilde{x}_+]=\int_{\hbar\beta/2}^{\hbar\beta/4} ids & \Big\{ \frac m 2 
 \left(\frac{dx_+(t-is+i\hbar\beta/2)}{ids}\right)^2  \nonumber\\
 & -V(x_+(t-is+i\hbar\beta/2))\Big\} . \label{S+1} 
\end{align}
After changing the integration variable from $s$ to $-s+\hbar\beta/2$,
we find that this is equal to $S_{3}^+[x_+]$. 
In a similar fashion, since $z=t+is$ with $0\le s\le \hbar\beta/4$ on 
$\mathcal{C}_{3}^+$, we can write
\begin{align}
 S_3^+[\tilde{x}_+]=\int_{\hbar\beta/4}^{0} ids &\Big\{ \frac m 2 
 \left(\frac{dx_+(-is+i\hbar\beta/2)}{ids}\right)^2  \nonumber   \\
 & -V(x_+(-is+i\hbar\beta/2))\Big\}. \label{S+3}
\end{align}
Again changing the variable from $s$ to $-s+\hbar\beta/2$ gives
$S_{3}^+[\tilde{x}_+]=S_{1}^+[x_+]$.
On $\mathcal{C}_{2}^+$, $z=s+i\hbar\beta/4$ with $0\le s \le t$, we have $t-z+i\hbar\beta/2=t-s+i\hbar\beta/4$.
Therefore, if we change the integration variable from $s$ to $t-s$, we can easily see that
$S_{2}^+[\tilde{x}_+]=S_{2}^+[x_+]$.  Combining all three components, we have $S_+[\tilde{x}_+]=S_+[x_+]$.
For $\mathcal{C}^-$, a similar relation holds. We have shown that the actions are invariant under the 
equilibrium transformation, 
 \begin{equation}
S_{\pm}[\tilde{x}_{\pm}]=S_{\pm}[{x}_{\pm}].
\end{equation}
 
For the influence functional, it is given by the double integrals and there are obviously more terms to deal with.
Nevertheless, we can apply similar change of integration variables and show that
\begin{equation}
\Psi[\tilde{x}_+,\tilde{x}_-]=\Psi[{x}_+,{x}_-].
\label{psitrans}
\end{equation}
The detailed derivation is presented in Appendix \ref{sec:appB}. We have shown that in equilibrium 
Eq.~(\ref{trans}) is a symmetry that the action in the path integral formalism of quantum Brownian motion 
satisfies.

\subsection{Equilibrium Fluctuation Dissipation Relations}

\begin{figure}
\includegraphics[width=0.4\textwidth]{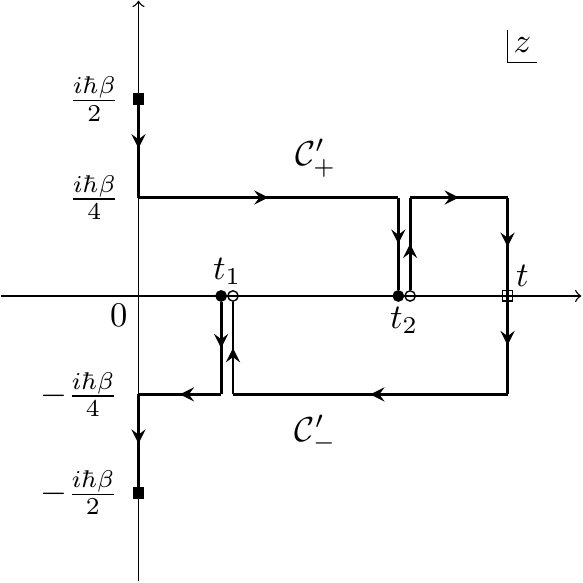}
\caption{Contours $\mathcal{C}'_\pm$ for $S^{\prime}_{\pm}[x_\pm]$ in Eq.~(\ref{actionprime}) for
the two-time correlation function $\langle A_{\mathrm S}(t_1) B_{\mathrm S}(t_2)\rangle$.
The open and filled circles indicate the endpoints where the system operators are inserted. 
The open and filled squares
are the endpoints which are traced over, respectively.}
\label{fig_corr}
\end{figure}

In this subsection, we look at the consequences of this symmetry in equilibrium. 
We consider the two-time correlation function, Eq.~(\ref{twotime}),
between two {\it system} operators $A_{\mathrm S}$ and $B_{\mathrm S}$,
and express it in the path integral representation on a deformed contour. 
Following the discussion in Sec.~\ref{sec:path}, we need to evaluate
an expression similar to Eq.~(\ref{one_point}). 
There are two instances where the two operators are inserted.
At time slice $t_2$, we have to insert $B_{\rm S}$ and evaluate the matrix element,
\begin{widetext}
\begin{align}
\langle x'_2,\bm{q}'_2 \vert e^{-i\theta_+(t_2) H_{\rm tot}/\hbar} 
B_{\mathrm{S}} e^{i\theta_+(t_2) H_{\rm tot}/\hbar} \vert 
x_2,\bm{q}_2\rangle&=\int d\bar{x}\int d\bar{x}'\int d\bar{\bm{q}}\; \langle \bar{x}'\vert B_{\mathrm{S}} 
\vert \bar{x}  \rangle \langle x'_2,\bm{q}'_2\vert e^{\beta H_{\rm tot}/4} 
\vert \bar{x}',\bar{\bm{q}}\rangle 
 \langle \bar{x},\bar{\bm{q}}\vert e^{-\beta H_{\rm tot}/4} \vert
x_2,\bm{q}_2\rangle \nonumber \\
&=\delta(\bm{q}_2-\bm{q}'_2)\int d\bar{x}\int d\bar{x}' \; 
\langle \bar{x}'\vert B_{\mathrm{S}} 
\vert \bar{x}  \rangle \langle x'_2\vert e^{\beta H_{\mathrm{S}}/4} 
\vert \bar{x}'\rangle  \langle \bar{x}\vert e^{-\beta H_{\mathrm{S}}/4} \vert
x_2\rangle\nonumber \\
&=\delta(\bm{q}_2-\bm{q}'_2)\langle x'_2\vert e^{\beta H_{\mathrm S}/4} B_{\mathrm{S}} 
e^{-\beta H_{\mathrm{S}}/4} \vert 
x_2\rangle . \label{corrB}
\end{align}
\end{widetext}
Going from the first to the second line in the above equation, we have represented the matrix elements as path integrals over
the paths along the imaginary time axis. We then performed path integrals over the bath variables 
and integrated over $\bar{\bm{q}}$
to obtain the delta function. This is again possible since the total Hamiltonian is Gaussian in the bath variable.
The details of this calculation is presented in Appendix \ref{sec:appC}.
We can now represent the last line of Eq.~(\ref{corrB}) in terms of the path integral 
over the system variable $x(z)$ where $z$ runs from $t_2+i\hbar\beta/4$ to $t_2$ and then
comes back to $t_2+i\hbar\beta/4$ in the complex
time domain (see Fig.~\ref{fig_corr}). The matrix element of $B_{\mathrm S}$ is inserted
at $t_2$. 
At time slice $t_1$, $A_{\mathrm S}$ is inserted as
$\langle x'_1,\bm{q}'_1 \vert e^{-i \theta_-(t_1)H(t_1)/\hbar}A_{\rm S} 
e^{i \theta_-(t_1)H(t_1)/\hbar}
\vert x_1,\bm{q}_1\rangle$, which results in after the similar calculation
\begin{equation}
\delta(\bm{q}_1-\bm{q}'_1)\langle x'_1\vert e^{-\beta H_{\mathrm{S}}/4} A_{\mathrm{S}} 
e^{\beta H_{\mathrm{S}}/4} \vert 
x_1\rangle. 
\label{corrA}
\end{equation}
Equation~(\ref{corrA}) is expressed in terms of the path integral over
$x(z)$ with $z$ now running back and forth between $t_1-i\hbar\beta/4$ and $t_1$. 

Apart from these two quantities, there are other matrix elements for the evaluation of
$\langle A_{\mathrm S}(t_1) B_{\mathrm S}(t_2)\rangle$. But these have already been evaluated for Eqs.~(\ref{trace}) 
and (\ref{one_point}). 
Combining all these, we have a path integral representation of the two-time correlation function as
\begin{align}
&\langle A_{\mathrm S}(t_1) B_{\mathrm S}(t_2)\rangle=\frac 1 {Z(0)}\int dx_i \int dx_f 
 \int d\bar{x}_b \int d\bar{x}'_b  \nonumber \\
 &\times  \int d\bar{x}_a \int d\bar{x}'_a  \; \langle \bar{x}'_b \vert B_{\mathrm S} \vert \bar{x}_b\rangle
 \langle \bar{x}'_a \vert A_{\mathrm S} \vert \bar{x}_a\rangle \nonumber \\
&\times\left(\int_{x_i}^{\bar{x}_b}+\int_{\bar{x}'_b}^{x_f}\right)\mathcal{D}x_+(z)
\left(\int_{x_f}^{\bar{x}'_a}+\int_{\bar{x}_a}^{x_i}\right)\mathcal{D}x_-(z) \nonumber \\
&\times  \exp\left(\frac{i}{\hbar}S'_+[x_+]+\frac{i}{\hbar}S'_-[x_-]-\frac{1}{\hbar}\Psi[x_+,x_-]\right),
 \label{corr_path}
\end{align}
where
\begin{equation}
S'_{\pm}[x_{\pm}]=\int_{\mathcal{C}'_{\pm}}dz\;\left[ \frac m 2 
\left(\frac{dx_{\pm}}{dz}\right)^2-V(x_{\pm}(z)) \right].
\label{actionprime}
\end{equation}
with the contours $\mathcal{C}^\prime_{\pm}$ shown in Fig.~\ref{fig_corr}. 
Compared to $\mathcal{C}_{\pm}$, the contours $\mathcal{C}^\prime_{\pm}$ have additional branches 
running along the imaginary time axis as mentioned above. In Eq.~(\ref{corr_path}),
the endpoints for the path integrals for $x_{+}$ indicate the constraints, 
$x_+(i\hbar\beta/2)=x_i$,
$x_+(t^-_2)=\bar{x}_b$, $x_+(t^+_2)=\bar{x}^\prime_b$ and $x_+(t)=x_f$. The endpoints
for $x_-(z)$ are similarly given as $x_-(-i\hbar\beta/2)=x_i$,
$x_-(t^-_1)=\bar{x}_a$, $x_-(t^+_1)=\bar{x}^\prime_a$ and $x_-(t)=x_f$. 
As we can deduce from the delta function for the bath variable in 
Eqs.~(\ref{corrB}) and (\ref{corrA}), 
the system operator insertion does
not have any effect on the path integrals for the bath variables, which
can be integrated over in the exactly same way as before.
Therefore, $\Psi[x_+,x_-]$ in Eq.~(\ref{corr_path}) is again given by the same 
expression as in Eq.~(\ref{psi}), for which $x_{\pm}(z)$ do not have vertical branches
at $t_1$ and $t_2$.  

\begin{figure}
  \includegraphics[width=0.4\textwidth]{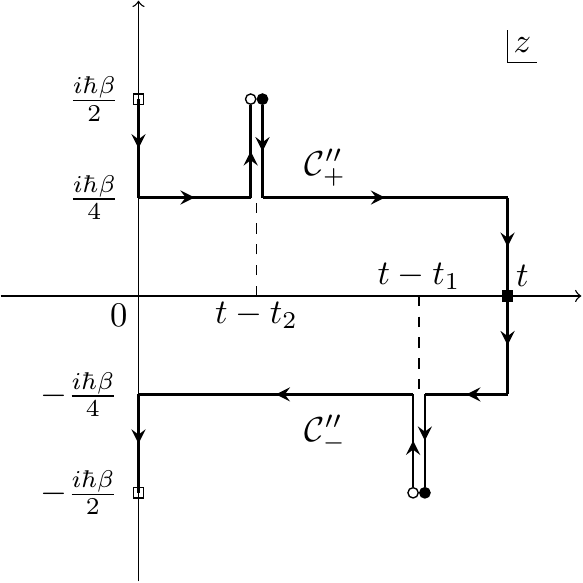}
\caption{Contours for $S^{\prime\prime}_{\pm}[x_\pm]$ which appears after taking the 
field transformations on the two-time correlation function. Note how the endpoints of the paths
indicated by open and filled circles and squares are changed from Fig.~\ref{fig_corr} after the transformation.}
  \label{fig_trans}
  \end{figure}

We now apply the field transformations given in 
Eq.~(\ref{trans}) to the path integral expression 
for the two-time correlation function in Eq.~(\ref{corr_path}). We first 
rewrite Eq.~(\ref{corr_path}) using $\tilde{x}_\pm$ and apply the transformations.
As before, $\Psi[\tilde{x}_+,\tilde{x}_-]=\Psi[x_+,x_-]$.
After applying the same integration variable change as before, we find that
that $S^\prime_{\pm} [ \tilde{x}_{\pm}]=S^{\prime\prime}_{\pm}[x_{\pm}]$,
where $S^{\prime\prime}_\pm$ is the same as Eq.~(\ref{actionprime}) except that the 
integral is now over the contour $\mathcal{C}^{\prime\prime}_{\pm}$ shown in 
Fig.~\ref{fig_trans}. On $\mathcal{C}^{\prime\prime}_+$, the path $x_+(z)$ now has two parts; one that runs from 
$x_+(i\hbar\beta/2)=x_f$ to $x_+(t-t_2+i\hbar\beta/2)=\bar{x}^\prime_b$
and the other from $x_+(t-t_2+i\hbar\beta/2)=\bar{x}_b$ to
$x_+(t)=x_i$. Similarly, for $x_-(z)$ on $\mathcal{C}^{\prime\prime}_-$, we have one
from $x_-(t)=x_i$ to $x_-(t-t_1-i\hbar\beta/2)=\bar{x}_a$,
and the other from $x_-(t-t_1-i\hbar\beta/2)=\bar{x}^\prime_a$ 
to $x_-(-i\hbar\beta/2)=x_f$. As can be seen from Figs.~\ref{fig_corr} and \ref{fig_trans},
$x_i$ and $x_f$ switch their places, but since these variables are integrated over, 
it does not make a difference in the calculation of the correlation function. 
On the other hand, the endpoints on which the operators are
inserted get interchanged (see how the open and filled circles in Fig.~\ref{fig_trans} are changed from those
in Fig.~\ref{fig_corr}).
We note that the path integral measure does not change, 
$\mathcal{D}\tilde{x}_+\mathcal{D}\tilde{x}_-
=\mathcal{D}x_+\mathcal{D}x_-$ with appropriate changes of the endpoints.

The above consideration shows that, after the transformation, the path integral expression becomes 
the one for the two-time correlation function for two operators, 
$e^{-\beta H_{\rm tot}/2} \tilde{B}_{\mathrm S} e^{\beta H_{\rm tot}/2}$
inserted at time $t-t_2$ and  $e^{\beta H_{\rm tot}/2} \tilde{A}_{\mathrm S} e^{-\beta H_{\rm tot}/2}$ 
at time $t-t_1$, where
$\tilde{A}_{\rm S}$ is defined by
$\langle x\vert \tilde{A}_{\mathrm S} \vert x'\rangle=\langle x'\vert A_{\mathrm S}\vert x\rangle$ for 
all $\vert x\rangle$ and $\vert x'\rangle$. The operator 
$\tilde{B}_{\mathrm S}$ is similarly defined. 
Here the factors of $e^{\pm \beta H_{\rm tot}/2}$ account
for the upward vertical branches in $\mathcal{C}^{\prime\prime}_\pm$ shown in Fig.~\ref{fig_trans}.
We can therefore write that
\begin{equation}
\langle A_{\mathrm S}(t_1) B_{\mathrm S}(t_2)\rangle 
=\langle\tilde{A}_{\mathrm S}(t-t_1-\frac{i\hbar\beta}2)\tilde{B}_{\mathrm S}(t-t_2+\frac{i\hbar\beta}2 )
\rangle,
\label{fdr}
\end{equation}
where we have defined $\tilde{A}_{\mathrm S}(z)=e^{iH_{\rm tot}z/\hbar}\tilde{A}_{\mathrm S} e^{-iH_{\rm tot}z/\hbar}$.
This equation is the consequence of the transformation, Eq.~(\ref{trans}) for the two-time correlation function. 
We note that $\tilde{A}_{\mathrm S}$ is in fact an operator obtained from the time reversal transformation
of $A_{\mathrm S}$ \cite{sakurai}. This follows from the following consideration.
For an observable $O$, $\tilde{O}=\Theta O \Theta^{-1}$ and $\vert \tilde{x}\rangle=\Theta \vert x\rangle$
for the time reversal transformation $\Theta$. We then 
have $\langle \tilde{x} \vert \tilde{O} \vert \tilde{x}'\rangle=\langle x \vert O \vert x'\rangle^*$ \cite{sakurai}.
Since $\vert \tilde{x}\rangle =\vert x \rangle$ and $O$ is a hermitian operator, we have
$\langle x \vert \tilde{O} \vert x'\rangle=\langle x' \vert O \vert x\rangle$.
As noted in Ref.~\cite{sieb}, Eq.~(\ref{fdr}) can be interpreted as the invariance of the two time correlation function 
under the time reversal transformation
combined with the Kubo-Martin-Schwinger (KMS) relation \cite{kms,kms1} which holds in equilibrium.
We note that the KMS condition reads in our case 
$\langle A_{\rm S} (t_1) B_{\rm S}(t_2)\rangle=\langle B_{\rm S}(t_2-i\hbar\beta/2) A_{\rm S} (t_1+i\hbar\beta/2)\rangle$.
Equation (\ref{fdr}), in our case,
constitutes the equilibrium FDR for the quantum Brownian motion. 

We can rewrite the relation, Eq.~(\ref{fdr}), 
in terms of more familiar Green's functions, $iG^>(t_1,t_2)=\langle x(t_1) x(t_2) \rangle$ and 
$iG^<(t_1,t_2)=\langle x(t_2) x(t_1) \rangle$, as
\begin{equation}
 G^> (t_1,t_2) = G^< (t-t_2+\frac{i\hbar\beta}2, t-t_1-\frac{i\hbar\beta}2).
\end{equation}
Since we expect Green's functions depend only on $\tau\equiv t_1-t_2$, we have 
\begin{equation}
 G^> (\tau) = G^< (\tau+i\hbar\beta).
\label{fdr1}
 \end{equation}
In terms of the retarded $G^{\rm R}$ and advanced $G^{\rm A}$ Green's functions, we have
$G^{\rm R}(\tau)-G^{\rm A}(\tau)=G^>(\tau)-G^<(\tau)$. The Keldysh Green's function is given as
$G^{\rm K}(\tau)=\frac 1 2 [G^>(\tau)+G^<(\tau)]$. From Eq.~(\ref{fdr1}), the FDR takes the following form.
\begin{equation}
 \cosh\left(i\frac {\hbar\beta} 2\partial_\tau\right)[G^{\rm R}(\tau)-G^{\rm A}(\tau)] =
 2\sinh\left(i\frac {\hbar\beta} 2 \partial_\tau\right)G^{\rm K}(\tau).
\end{equation}

\section{non-equilibrium}
\label{sec:noneq}

In this section, we consider the case where the system Hamiltonian $H_{\mathrm S}(s)$
depends on time $s$ explicitly through the potential energy
$V(x;\lambda_s)$ with the time-dependent protocol $\lambda_s$ for $0\le s \le t$. 
As before, we develop the path integral formulation for this case. 
For the forward path in Eq.~(\ref{trace}), 
we have to evaluate the matrix element similar to Eq.~(\ref{mat_elem}) 
between the time slices
$s_k$ and $s_{k+1}$, which arises from the discretization of 
$U_{t,0}$. This is in the form of 
$\langle x_{k+1},\bm{q}_{k+1}\vert M^+_k \vert x_k,\bm{q}_k\rangle$, where
\begin{widetext}
  \begin{align}
M^+_k=&e^{-\frac i\hbar \theta_+(s_{k+1})H_{\rm tot}(s_{k+1})}e^{-\frac i\hbar H_{\rm tot}(s_k) ds} 
e^{\frac i\hbar \theta_+(s_{k})H_{\rm tot}(s_k)}  \nonumber \\
\simeq& 1-\frac{i}{\hbar} ds (1+\dot{\theta}_+(s_k)) H_{\rm tot}(s_k) -\frac{i}{\hbar} ds \theta_+(s_k) 
 \int_0^1 d\xi\;
e^{-\frac i\hbar \xi \theta_+(s_k) H_{\rm tot}(s_k)} (\partial_s H_{\rm tot}(s_k))
e^{\frac i\hbar \xi \theta_+(s_k) H_{\rm tot}(s_k)},  \label{time1}
\end{align}
\end{widetext}
where we have used an identity 
 \begin{equation}
   \frac{d}{dt}e^{O(t)}
  =\int_0^1 d\xi\; e^{\xi O(t)} \frac{dO(t)}{dt}
  e^{(1-\xi)O(t)}
 \end{equation}
valid for an arbitrary time-dependent operator $O(t)$.
If we only had the first two terms in Eq.~(\ref{time1}), the situation would be quite similar to 
that in the previous section for the time-independent case.  The matrix element would just give 
$\exp(i dz_+ \mathcal{L}^+_{\rm tot}/\hbar)$
where the Lagrangian $\mathcal{L}^+_{\rm tot}$ has again three components given by
Eqs.~(\ref{LS}), (\ref{LB}) and (\ref{LI}) except that the system Lagrangian $\mathcal{L}^+_{\mathrm S}$ has an
extra dependence on the external protocol $\lambda$. But as we will see below, because of the 
last term in Eq.~(\ref{time1}), the system Lagrangian gets significantly modified. 

The effect of the time-dependent Hamiltonian is in the last term in Eq.~(\ref{time1}) with 
$\partial_s H_{\rm tot}=\partial_s H_{\mathrm S}=\dot{\lambda_s}\partial_\lambda V(x;\lambda)$.
which we now evaluate for $\theta_+(s)=i\hbar\beta/4$.
The matrix element of this term is given by
\begin{align}
& -\frac{i}{\hbar} ds \delta(\bm{q}_{k+1}-\bm{q}_k)  
 \left(\frac{i\hbar\beta}4\right) \int_0^1 d\xi\;
\langle x_{k+1}| e^{\frac 1 4 \xi \beta H_{\rm S}(s_k)} \nonumber \\
&\quad\quad\quad \times (\partial_s H_{\mathrm S}(s_k))
e^{-\frac 1 4 \xi \beta H_{\mathrm S}(s_k)} |x_k\rangle  ,
\end{align}
where we have again performed the path integral over the bath variables as in Eq.~(\ref{corrB}) 
following the procedures outlined in Appendix \ref{sec:appC} to get the delta function in $\bm{q}$. 
Note that the matrix element now is with respect to the system variable only.
Therefore this term modifies the system Lagrangian $\mathcal{L}_{\mathrm S}(x_+(z))$  when $z$ is
on the contour $\mathcal{C}^+_2$ in Fig.~\ref{fig_one_point}. 
On the other parts of the contour $\mathcal{C}^+$, where $\partial_s H_{\mathrm S}=0$,
the system Lagrangians are the same as the equilibrium ones. Therefore, on $\mathcal{C}^+_2$, we have a modified 
Lagrangian $\widehat{\mathcal{L}}^+_{\mathrm S}$ at time slice $s_k$ defined by
\begin{equation}
e^{\frac i\hbar ds \widehat{\mathcal{L}}^+_{\mathrm S}} 
= \langle x_{k+1}\vert 1-\frac i\hbar ds 
(H_{\mathrm S}(s_k) +F_+(s_k)) \vert x_k\rangle,
\label{L+}
\end{equation}
where 
\begin{equation}
 F_+(s)\equiv i \frac{\beta\hbar}{4} \int_0^1 d\xi\; 
 e^{\frac 1 4 \xi\beta H_{\mathrm S}(s)}\partial_s H_{\mathrm S}(s)
 e^{-\frac 1 4 \xi\beta 
 H_{\mathrm S}(s)}.
 \label{G+}
\end{equation}

For the backward path, we have to evaluate $\langle x_{k},\bm{q}_{k}\vert M^-_k \vert x_{k+1},\bm{q}_{k+1}\rangle$,
where
\begin{equation}
M^-_k=e^{-\frac i\hbar \theta_-(s_{k})H_{\rm tot}(s_{k})}e^{\frac i\hbar H_{\rm tot}(s_k) ds} 
e^{\frac i\hbar \theta_-(s_{k+1})H_{\rm tot}(s_{k+1})}.
\end{equation}
Following the same argument, we find that the time-dependent Hamiltonian modifies the system Lagrangian 
when the path $x_-(z)$ is on
the contour $\mathcal{C}^-_2$ in Fig.~\ref{fig_one_point}. The modified system Lagrangian   
$\widehat{\mathcal{L}}^-_{\mathrm S}$ at time slice $s_k$ is defined by
\begin{equation}
e^{-\frac i\hbar ds \widehat{\mathcal{L}}^-_{\mathrm S} } 
= \langle x_{k}\vert 1+\frac i\hbar ds 
(H_{\mathrm S}(s_k) +F_-(s_k)) \vert x_{k+1}\rangle,
\label{L-}
\end{equation}
where 
\begin{equation}
 F_-(s)\equiv -i \frac{\beta\hbar}{4} \int_0^1 d\xi\; 
 e^{-\frac 1 4 \xi\beta H_{\mathrm S}(s)}\partial_s H_{\mathrm S}(s)
 e^{\frac 1 4 \xi\beta 
 H_{\mathrm S}(s)}.
 \label{G-}
\end{equation}
We can easily see that $ \widehat{\mathcal{L}}^-_{\mathrm S}= (\widehat{\mathcal{L}}^+_{\mathrm S})^*$.
  
Since there is no change in the bath part of the action even in the case of the time-dependent Hamiltonian,
we get the same influence functional $\Psi$ as in Eq.~(\ref{psi}). Therefore, the only change 
that the time-dependent Hamiltonian makes, for example, in the 
the normalization condition, Eq.~(\ref{S_ave1}) is in the system action $S_\pm$ given in Eq.~(\ref{action}) on the 
contours $\mathcal{C}^\pm_2$. We therefore have, instead of $S_\pm$, 
a new expression for the system action as
\begin{equation}
\widehat{S}_{\pm}[x_\pm;\lambda]=\widehat{S}^{\pm}_1[x_\pm;\lambda_0]+
\widehat{S}^{\pm}_3[x_\pm;\lambda_t] + \widehat{S}_2^\pm[x_\pm;\lambda],
\end{equation}
where
\begin{align}
 &\widehat{S}^{\pm}_1[x_\pm;\lambda_0]=\int_{\mathcal C^\pm_1}dz\; 
 \mathcal{L}^\pm_{\mathrm S}(x_\pm (z),\dot{x}_\pm(z);\lambda_0) 
 \label{action1}\\
 &\widehat{S}^{\pm}_3[x_\pm;\lambda_t]=\int_{\mathcal C^\pm_3} dz\; 
 \mathcal{L}^\pm_{\mathrm S}(x_\pm (z),\dot{x}_\pm(z);\lambda_t) 
 \label{action3}
\end{align}
are given in terms of the original Lagrangians in Eq.~(\ref{LS})
with the dependence of 
$V$ on the protocol $\lambda$ inserted. On the other hand, we have 
\begin{align}
&\widehat{S}^\pm_2[x_\pm;\lambda] \nonumber \\
&=\pm\int_0^t ds\; \widehat{\mathcal{L}}^\pm_{\mathrm S} 
(x_\pm (s\pm \frac {i\hbar\beta} 4 ),\dot{x}_\pm (s\pm \frac {i\hbar\beta} 4 );\lambda_{s})
\label{actionA}
\end{align}
with the modified Lagrangian
$\widehat{\mathcal{L}}^\pm_{\mathrm S}$ determined from Eqs.~(\ref{L+}) and (\ref{L-}).

\subsection{Quantum Fluctuation Theorem}

Here we show that a quantum fluctuation theorem can be obtained by studying the behavior of the actions 
under the transformation, Eq.~(\ref{trans}).
Specifically we look at the normalization condition, Eq~(\ref{S_ave1}), which 
can be written in the presence of the time-dependent protocol as
\begin{align}
1=&\frac 1 {Z(0)}\int dx_i \int dx_f
\int_{x_i}^{x_f}\mathcal{D}x_+
\int_{x_f}^{x_i}\mathcal{D}x_-  \label{S_ave2} \\
&\times  e^{ \frac{i}{\hbar}\widehat{S}_+[x_+;\lambda]+\frac{i}{\hbar}\widehat{S}_-[x_-;\lambda]-\frac{1}{\hbar}\Psi[x_+,x_-]},
\nonumber
\end{align}
By the same methods used in Eqs.~(\ref{S+1}) and (\ref{S+3}), 
we find that,
under the transformation, Eq.~(\ref{trans}),
\begin{align}
 &\widehat{S}^{\pm}_1[\tilde{x}_\pm;\lambda_0] =\widehat{S}^{\pm}_3[x_\pm;\lambda_0]
 \equiv\widehat{S}^{\pm}_3[x_\pm;\tilde{\lambda}_t] \label{s1trans}\\
 &\widehat{S}^{\pm}_3[\tilde{x}_\pm;\lambda_t] =\widehat{S}^{\pm}_1[x_\pm;\lambda_t]
 \equiv\widehat{S}^{\pm}_1[x_\pm;\tilde{\lambda}_0], \label{s3trans}
\end{align}
where we defined the time-reversed protocol, $\tilde{\lambda}_s\equiv \lambda_{t-s}$.
Since $\widehat{\mathcal{L}}^\pm_{\mathrm S}$  
in general has a different form from $\mathcal{L}^\pm_{\mathrm S}$,
the action $\widehat{S}_2^{\pm}$ does not show simple transformations
as $S^\pm_1$ or $S^\pm_3$ does. In fact, if we consider $\widehat{S}_2^{\pm}[\tilde{x}_{\pm};\lambda]$
in Eq.~(\ref{actionA}), we find,
after changing the time integration variable from $s$ to $t-s$,
\begin{align}
&\widehat{S}^\pm_2[\tilde{x}_\pm;\lambda] \nonumber \\
&=\int_0^t ds\; \widehat{\mathcal{L}}^\pm_{\mathrm S} 
(x_\pm (s\pm \frac {i\hbar\beta} 4 ),-\dot{x}_\pm (s\pm \frac {i\hbar\beta} 4 );\tilde{\lambda}_{s}).
\end{align}
Since the Lagrangian $\widehat{\mathcal{L}}^\pm_{\mathrm S}$ is not guaranteed to have a standard form 
with a velocity squared term, it is in general different from $\widehat{S}_2^{\pm}[x_{\pm};\tilde{\lambda}]$.
We define their difference as 
\begin{equation}
 \widehat{S}_2^{\pm}[\tilde{x}_{\pm};\lambda]=
 \widehat{S}_2^{\pm}[x_{\pm};\tilde{\lambda}]\pm \Sigma_{\pm}[x_{\pm};\tilde{\lambda}],
 \label{sigmadef}
\end{equation}
where the actual form of $\Sigma_\pm$ can be calculated once $F_\pm$ in Eqs.~(\ref{G+})
and (\ref{G-}) are known. We can easily derive, regardless of its actual form, $\Sigma_\pm$ satisfies
\begin{equation}
\Sigma_{\pm}[x_{\pm};\tilde{\lambda}]=-\Sigma_{\pm}[\tilde{x}_{\pm};\lambda].\label{sigmarel}
\end{equation}
This term signals
the breaking of symmetry in the non-equilibrium situation which was satisfied in the equilibrium case.
As we will see, it also plays an important role in the fluctuation theorem we are about to derive. 

We now derive the fluctuation theorem.
Let us consider
\begin{equation}
\left\langle e^{\frac i\hbar \left( \Sigma_+[x_+;\lambda]+\Sigma_-[x_-;\lambda]\right)}\right\rangle 
\end{equation}
where the average $\langle\cdots\rangle$ is done with respect to the action in the integrand of Eq.~(\ref{S_ave2}).  
We then change the path integral variables from 
$x_{\pm}$ to $\tilde{x}_{\pm}$.  
Using Eqs.~(\ref{s1trans}), (\ref{s3trans}), (\ref{psitrans}), (\ref{sigmadef}),(\ref{sigmarel})
and the invariance of the Jacobian of the transformation, we find 
\begin{align}
&\left\langle e^{\frac i\hbar \left( \Sigma_+[x_+;\lambda]+\Sigma_-[x_-;\lambda]\right)}\right\rangle 
=\frac 1 {Z(0)}\int dx_i \int dx_f \label{ft1} \\
&\times \int_{x_f}^{x_i}\mathcal{D}x_+
\int_{x_i}^{x_f}\mathcal{D}x_-  
e^{\frac{i}{\hbar}\widehat{S}_+[x_+;\tilde{\lambda}]+\frac{i}{\hbar}\widehat{S}_-[x_-;\tilde{\lambda}]
-\frac{1}{\hbar}\Psi[x_+,x_-]}.
\nonumber
\end{align}
The right hand side is just the normalization with the reverse protocol $\tilde{\lambda}$ 
except for the factor of $1/Z(0)$. 
We therefore have
\begin{align}
\left\langle e^{\frac i\hbar \left( \Sigma_+[x_+;\lambda]+\Sigma_-[x_-;\lambda]\right)}
\right\rangle &=\frac{Z(t)}{Z(0)}
 \\
&\equiv
e^{ -\beta \left( \mathcal{F}(t)-\mathcal{F}(0)\right)}, \nonumber
\end{align}
where we have defined the free energy $\mathcal{F}\equiv -(1/\beta)\ln Z$.
If we identify
\begin{equation}
\frac i\hbar \left( \Sigma_+[x_+;\lambda]+\Sigma_-[x_-,\lambda]\right)\equiv-\beta\Xi
\label{workdef}
\end{equation}
with $\Xi$ being the quantum mechanical work-like quantity defined on the fluctuating trajectory, we have
the Jarzynski-like fluctuation theorem,
\begin{equation}
\left\langle e^{-\beta \Xi}\right\rangle =e^{-\beta\Delta\mathcal{F}}.
\end{equation}
The quantity defined in Eq.~(\ref{workdef}) has some similarity to the work functional that appears in the path integral representation of the 
work statistics for the TPM scheme \cite{funo}. Both reduce 
to the familiar classical expression for work in the classical limit as we shall show for Eq.~(\ref{workdef}) below. 
But since our formalism is essentially independent of measurement, they are fundamentally different quantities. 
It is conceptually more similar to the quantum work not based on the TPM scheme as defined in Refs.~\cite{solinas1,miller,hofer,sampaio}.
In those works, quantum trajectories are also considered in the form of the path integral under the constraint that 
observables have specified values \cite{solinas1}, the trajectories consisting of the projectors of power operators \cite{miller},
the Keldysh quasi-probability distribution \cite{hofer}, and the Bohmian trajectories \cite{sampaio}.    
But the work-like quantity we find in Eq.~(\ref{workdef}) is different from these in that its average value 
for a closed system \cite{abc} is not equal to 
the change in the average internal energy.  Equation (\ref{workdef}), arising from the
symmetry properties of the SK path integrals, describes
an intrinsic property of a quantum process as in Refs.~\cite{miller,hofer} where the quasi-probability distribution
arises, and in addition satisfies the Jarzynski-type equality and reduces to the classical work expression
in the classical limit. 
  
To the lowest order in $\hbar$, $F_{\pm}$ in Eqs.~(\ref{G+}) and (\ref{G-}) becomes
\begin{equation}
F_{\pm}(s)\simeq \pm \frac{i\hbar\beta}4\dot{\lambda}_s\partial_\lambda V.
\end{equation}
Then $\widehat{S}^\pm_2$ can be obtained from Eq.~(\ref{L+}) and (\ref{L-}) as
\begin{align}
   &\widehat{S}^\pm_2[x_\pm;\lambda] \simeq
   \pm\int_0^t ds \; \Big[ \frac{m}{2} 
   \Big\{ \frac{d}{ds} x_\pm(s\pm \frac{i\hbar\beta}4) 
    \Big\}^2 \label{action_cl} \\
&  -V(x_\pm(s\pm\frac{i\hbar\beta}4);\lambda_s) \mp \frac{i\hbar\beta}4\dot{\lambda}_s\partial_\lambda 
V(x_\pm(s\pm\frac{i\hbar\beta}4);\lambda_s)\Big].  \nonumber 
  \end{align}
In order to calculate $\Sigma_\pm$ defined in Eq.~(\ref{sigmadef}), we insert $\tilde{x}_\pm$ into 
$\widehat{S}^\pm_2$ and see how the action changes from the original one. Using Eq.~(\ref{trans}) and changing
the integration variable from $s$ to $t-s$, we have
\begin{align}
   &\widehat{S}^\pm_2[\tilde{x}_\pm;\lambda] \simeq
   \pm\int_0^t ds \; \Big[ \frac{m}{2} 
   \Big\{ \frac{d}{ds} x_\pm(s\pm \frac{i\hbar\beta}4) 
    \Big\}^2 \label{action_cl1} \\
&  -V(x_\pm(s\pm\frac{i\hbar\beta}4);\tilde{\lambda}_s) \pm 
\frac{i\hbar\beta}4\dot{\tilde{\lambda}}_s\partial_{\tilde{\lambda}} 
V(x_\pm(s\pm\frac{i\hbar\beta}4);\tilde{\lambda}_s)\Big], \nonumber 
  \end{align}
where we have used the fact that under the change of variable $s\to t-s$, $\dot{\lambda}_s\to -d\lambda_{t-s}/ds=
-\dot{\tilde{\lambda}}_s$ with $\tilde{\lambda}_s\equiv\lambda_{t-s}$. 
This expression differs from $\widehat{S}^\pm_2[x_\pm,\tilde{\lambda}]$ by the change of sign in the last term.
From Eq.~(\ref{sigmadef}), we deduce that
\begin{equation}
 \Sigma_\pm[x_\pm,\lambda]\simeq \frac{i\hbar\beta}{2}\int_0^tds\; \dot{\lambda}_s\partial_\lambda
 V(x_\pm(s\pm \frac{i\hbar\beta}4),\lambda_s).
\end{equation}
Therefore, to the leading order in $\hbar$, we have from Eq.~(\ref{workdef}) 
\begin{align}
 \Xi \simeq \frac 1 2 \int_0^t ds\; \dot{\lambda_s}   \partial_\lambda \Big[V(x_+(s);\lambda_s) 
 + V(x_-(s);\lambda_s) \Big] . \label{work_cl} 
\end{align}
We will show in the next section that this expression further reduces to the familiar expression for
the classical work $\mathcal{W}_c=\int_0^t ds\; \dot{\lambda}_s \partial_\lambda V(r(s);\lambda_s)$
defined along the classical stochastic path $r(s)$.

\subsection{Example: a pulled harmonic oscillator}
In this subsection, we consider a concrete example to elucidate the nature of
quantum work and FT discussed above. As an exactly solvable model,
the driven harmonic oscillator
has been studied in various settings for the calculation of 
the distribution function of the quantum
work based on the TPM scheme. See, for example, Refs.~\cite{dho1,dho2,dho3}.
In this paper, we consider a harmonic oscillator with its center
moving with a specified protocol $\lambda_s$ for $0\le s\le t$ interacting
with the bath. The system Hamiltonian is given by
\begin{equation}
 H_{\rm S}=\frac{p^2}{2m}+ \frac 1 2 m\omega^2_0 (x-\lambda_s)^2
\end{equation}
with $ \partial_s H_{\rm S} =-m\omega^2_0 \dot{\lambda}_s (x-\lambda_s)$.
Using 
\begin{equation}
e^{-a H_{\rm S}}x e^{a H_{\rm S}}=x\cosh(\hbar\omega_0 a)-i\frac{p}{m\omega_0}\sinh(\hbar\omega_0 a),
\end{equation}
we have from Eqs.~(\ref{G+}) and (\ref{G-})
\begin{align}
F_\pm(s)=&\mp i m\omega_0
 \dot{\lambda}_s\Big[ 
 (x-\lambda_s)\sinh(\beta\hbar\omega_0/ 4) \nonumber \\ 
&~~~~~~~~~~~~~~ \mp i \frac{p}{m\omega_0}\{\cosh(\beta\hbar\omega_0/4)-1\}
 \Big].
\end{align}
In order to calculate $\widehat{\mathcal{L}}^+_{\mathrm S}$ in Eq.~(\ref{L+}), we evaluate the matrix element
on the right hand side, which is given by
\begin{widetext}
\begin{align}
\langle x_{k+1} | e^{- i ds (H_{\rm S}(s_k)+F_+(s_k))/\hbar} 
|x_k\rangle 
&= \sqrt{\frac{m}{2\pi i\hbar ds}}\exp\left[-\frac{i}{\hbar} ds  \frac{m\omega^2_0}{2}(x_k-\lambda_{s_k})^2 
-\frac{ds} \hbar m\omega_0
 \dot{\lambda}_{s_k}
 (x_k-\lambda_{s_k})\sinh(\frac{\beta\hbar\omega_0}4 ) \right]\nonumber \\
 &\times 
 \exp\Big[ \frac i \hbar ds \frac{m}{2} \Big\{ (\frac{x_{k+1}-x_k}{ds}) 
+\dot{\lambda}_{s_k}( \cosh(\frac {\beta\hbar\omega_0} 4)-1 ) \Big\}^2\Big].
 \nonumber 
 \end{align}
 \end{widetext}
This is equal to $\exp[(i/\hbar)ds\widehat{\mathcal{L}}^+_{\mathrm S}]$ for this time slice. 
The Lagrangian $\widehat{\mathcal{L}}^-_{\mathrm S}$ for the backward contour can be obtained by the
complex conjugate of $\widehat{\mathcal{L}}^+_{\mathrm S}$. 
Collecting the contributions from all the time slices, we have the action $\widehat{S}^{\pm}_2$ on the forward and backward branches,
$\mathcal{C}^\pm_2$ in Fig.~\ref{fig_one_point}.
We have
\begin{align}
   &\widehat{S}^\pm_2[x_\pm;\lambda] \label{action_sho}\\
   =&\pm\int_0^t ds \; \Big[ \frac{m}{2} 
   \Big\{ \frac{d}{ds} x_\pm(s\pm \frac{i\hbar\beta}4) 
   +\dot{\lambda}_s ( \cosh(\frac {\beta\hbar\omega_0} 4)-1 ) \Big\}^2 \nonumber \\
&\quad\quad\quad\quad  -\frac{m\omega^2_0}{2}(x_\pm(s\pm\frac{i\hbar\beta}4)-\lambda_s)^2 \nonumber \\
&\quad\quad\quad\quad  \pm i m\omega_0\dot{\lambda}_s
 (x_\pm(s\pm \frac{i\hbar\beta}4)-\lambda_{s})\sinh(\frac {\beta\hbar\omega_0} 4 )\Big].  \nonumber 
  \end{align}
We note that the explicit time dependence in the system Hamiltonian is responsible for the two new
terms proportional to $\dot{\lambda}_s$.   

We now put $\tilde{x}_\pm$ instead of $x_\pm$ and see how the action changes from above. Upon changing of
the integration variable from $s$ to $t-s$, we have
\begin{align}
   &\widehat{S}^{\pm}_2[\tilde{x}_{\pm};\lambda] \label{action_sho1}\\ 
    =&\pm\int_0^t ds \; \Big[ \frac{m}{2} 
   \Big\{ \frac{d}{ds} x_\pm(s\pm \frac{i\hbar\beta}4) 
   +\dot{\tilde{\lambda}}_s ( \cosh(\frac {\beta\hbar\omega_0} 4)-1 ) \Big\}^2 \nonumber \\
&\quad\quad\quad\quad  -\frac{m\omega^2_0}{2}(x_\pm(s\pm\frac{i\hbar\beta}4)-\tilde{\lambda}_s)^2 \nonumber \\
&\quad\quad\quad\quad  \mp i m\omega_0\dot{\tilde{\lambda}}_s
 (x_\pm(s\pm \frac{i\hbar\beta}4)-\tilde{\lambda}_{s})\sinh(\frac {\beta\hbar\omega_0} 4 )\Big],  \nonumber 
 \end{align}
By comparing Eq.~(\ref{action_sho1}) with $\widehat{S}^\pm_2[x;\tilde{\lambda}]$ and using Eq.~(\ref{sigmadef}), we identify
\begin{align}
 &\Sigma_{\pm}[x_{\pm};\lambda]\\  
 =& -2i m\omega_0\int_0^t ds\;
 \dot{\lambda}_s
 (x_{\pm}(s\pm \frac{i\hbar\beta}4)-\lambda_{s}) \sinh(\frac {\beta\hbar\omega_0} 4 ) . \nonumber 
\end{align}
The quantum mechanical work-like quantity $\Xi$ defined on the trajectory, Eq.~(\ref{workdef}) is then given by
\begin{equation}
 \Xi=-\frac{4}{\beta\hbar\omega_0}\sinh(\frac {\beta\hbar\omega_0} 4 ) 
 \int_0^t ds\;m\omega^2_0 \dot{\lambda}_s
 (x_{c}(s)-{\lambda}_{s}), \label{work_sho}
\end{equation}
where
\begin{equation}
x_c(s)=\frac 1 2 [ x_{+}(s+ \frac{i\hbar\beta}4) + x_{-}(s- \frac{i\hbar\beta}4) ]
\label{xc}
\end{equation}
is the variable corresponding to the classical path to be explained in the next section.

\section{Classical Limit}
\label{sec:cl}

In this section, we take the classical limit and show that the above path integral formalism reduces 
to the MSRJD functional integral for the classical stochastic field satisfying a generalized Langevin equation.
We also show that the field transformations used in the previous sections become exactly those used
in classical stochastic thermodynamics
for the derivation of classical fluctuation theorems \cite{mallick,abc1}.

In the $\hbar\to 0$ limit, the path integral in Eq.~(\ref{S_ave2}) is dominated by the stationary points of the action 
$\widehat{S}_++\widehat{S}_-+i\Psi$. In order to study the fluctuation around the saddle point,
it is convenient to use \cite{kamenev,grabert,weiss} for $0\le s \le t$
the classical field $x_c$ defined in Eq.~(\ref{xc}) and the quantum field defined by
\begin{equation}
x_q(s)\equiv x_{+}(s+\frac{i\hbar\beta}4) - x_{-}(s-\frac{i\hbar\beta}4) .\label{xq}
\end{equation}
To the leading order of $O(\hbar)$, the solution to the stationary point equation for $x_q$ is just $x_q(s)=0$, while
$x_c$ satisfies a deterministic equation \cite{grabert,weiss}. 
Here we write the quantum mechanical path integral as a functional integral over a fluctuating variables,
$x_c(s)=O(1)$ and $x_q(s)=O(\hbar)$ \cite{weiss}. 
We first consider how $\widehat{S}_2^\pm$ in Eq.~(\ref{actionA}) behaves in the $\hbar\to 0$ limit. To the leading order in
$\hbar$, we have
\begin{align}
\widehat{S}^+_2+\widehat{S}^-_2\simeq \int_0^t ds &[-x_q(s)\{ m\ddot{x}_c(s)+\partial_{x_c}V(x_c(s);\lambda_s)\} \nonumber \\
&-\frac{i\hbar\beta}{2}\dot{\lambda}_s\partial_\lambda V(x_c(s);\lambda_s) ], \label{s2cl}
\end{align}
where we have used the boundary condition $x_q(0)=x_q(t)=0$. 

The connection with the classical MSRJD fields, denoted here by $r(s)$ and $\hat{x}(s)$, 
are made via \cite{weiss}
\begin{align}
& r(s)\equiv \frac 1 2 (x_+(s)+x_-(s)), \\
&\hat{x}(s)\equiv\lim_{\hbar\to 0} \frac 1 \hbar (x_+(s)-x_-(s)) .
\end{align} 
We note from Eqs.~(\ref{xc}) and (\ref{xq}) that
\begin{align}
& x_c(s)=r(s)+O(\hbar^2), \\
& x_q(s)=\hbar(\hat{x}(x)+\frac{i\beta}{2}\dot{r}(s))+O(\hbar^2).
 \label{xqxh}
\end{align}
Inserting these into Eq.~(\ref{s2cl}), we find that a part of the
integrand can be written as
a total time derivative of the system energy $\mathcal{H}_{\rm S}(s)
\equiv (m/2)\dot{r}^2(s)+V(r(s);\lambda_s)$ and we have
\begin{align}
& \lim_{\hbar\to 0}\frac{i}{\hbar}(\widehat{S}^+_2+\widehat{S}^-_2)=\frac{\beta}{2}(\mathcal{H}_{\rm S}(t)
-\mathcal{H}_{\rm S}(0))\nonumber \\
 &~~~~~~~~~~-i\int_0^t ds\; \hat{x}(s)\{ m\ddot{r}(s)+\partial_{r}V(r(s);\lambda_s)\} .
 \label{s2pm}
\end{align}
Now the actions given on the imaginary axis in Eqs.~(\ref{action1}) and (\ref{action3}) are given by
\begin{align}
 & \lim_{\hbar\to 0}\frac{i}{\hbar}(\widehat{S}^+_1+\widehat{S}^-_1)=-\frac{\beta}2\mathcal{H}_{\rm S}(0) \label{s1pm}\\
 & \lim_{\hbar\to 0}\frac{i}{\hbar}(\widehat{S}^+_3+\widehat{S}^-_3)=-\frac{\beta}{2}\mathcal{H}_{\rm S}(t). \label{s3pm}
\end{align}

The classical limit of the influence functional in Eq.~(\ref{psi}) can be calculated in a similar fashion.
We first note that for $0\le s\le t$, we can rewrite the kernel in Eq.~(\ref{kernel}) as
$K(s)=N(s)-\frac i 2 D(s)$ where
\begin{align}
& N(s)=\sum_n\frac{c^2_n}{2m_n\omega_n}\coth(\frac 1 2 \beta\hbar\omega_n)\cos(\omega_n s),
\\
& D(s)=\sum_n\frac{c^2_n}{m_n\omega_n}\sin(\omega_n s).
\end{align}
In the $\hbar\to 0$ limit, $N(s)\to (1/\hbar) (1/\beta)\gamma(s)$,
where
\begin{equation}
 \gamma(s)\equiv \sum_n \frac{c^2_n}{m_n\omega^2_n} \cos(\omega_n s).
\end{equation}
Inserting Eqs.~(\ref{xc}), (\ref{xq}) and (\ref{xqxh}) into Eq.~(\ref{psi}), we obtain
after a lengthy algebra
\begin{align}
\lim_{\hbar\to 0}\frac 1\hbar \Psi&=\frac 1 {2\beta} \int_0^t ds\int_0^t du\; \hat{x}(s)\gamma(s-u)\hat{x}(u)  \\
 & -i\int_0^t ds\;\hat{x}(s)\int_0^s du\;D(s-u)r(u) \nonumber \\
 & -ir(0)\int_0^t ds\; \hat{x}(s)\gamma(s)+i\mu\int_0^t ds\;\hat{x}(s)r(s). \nonumber 
\end{align}
Using the fact that
$D(s)=- d\gamma(s)/ds$ and $\mu=\gamma(0)$, and integrating by parts the second term, we finally have
\begin{align}
  \lim_{\hbar\to 0}\frac 1\hbar \Psi=& \frac 1 {2\beta} \int_0^t ds\int_0^t du\; \hat{x}(s)\gamma(s-u)\hat{x}(u) 
\nonumber \\
&+i \int_0^t ds\;\hat{x}(s)\int_0^s du\;\gamma(s-u)\dot{r}(u). \label{psihbar}
\end{align}
Combining Eqs.~(\ref{s2pm}), (\ref{s1pm}), (\ref{s3pm}) and (\ref{psihbar}), we find that the quantum path integral behaves 
as a path integral over $Z^{-1}(0)\int\mathcal{D}r(s)\int\mathcal{D}\hat{x}\;\exp[S_{\rm MSRJD}[r,\hat{x}]]$, 
where the MSRJD action is given by 
\begin{align}
S_{\rm MSRJD}=& - \frac 1 {2\beta} \int_0^t ds\int_0^t du\; \hat{x}(s)\gamma(s-u)\hat{x}(u) 
\nonumber \\
&-i \int_0^t ds\;\hat{x}(s)\int_0^s du\;\gamma(s-u)\dot{r}(u) \nonumber \\
&-i\int_0^t ds\; \hat{x}(s)\{ m\ddot{r}(s)+\partial_{r}V(r(s);\lambda_s)\} \nonumber \\
& -\beta\mathcal{H}_{\rm S}(0).
\end{align}
This is exactly the MSRJD action for the generalized Langevin equation 
\begin{equation}
m\ddot{r}(s)+\partial_{r} V(r;\lambda_s) +\int_0^s du\; \gamma(s-u)\dot{r}(u)=\xi(s),
\end{equation}
where the noise $\xi(s)$ satisfies
\begin{equation}
\langle \xi(s)\rangle=0,~~~~~~~\langle\xi(s)\xi(s')\rangle=\frac{1}{\beta}\gamma(s-s').
\end{equation}
We note that for the case of a factorized initial state of system and bath,
a term which depends on the initial value of $r(s)$ appears in the classical
limit of the quantum Brownian motion \cite{weiss}. This does not
show here for the initial state given by Eq.~(\ref{init}) \cite{grabert}.
The quantum mechanical work-like quantity, which we found in Eq.~(\ref{workdef}) becomes 
in the classical limit, Eq.~(\ref{work_cl}).  In terms of the MSRJD variables, it reduces to the familiar expression
\begin{equation}
\mathcal{W}_c=\int_0^t ds\; \dot{\lambda}_s \frac{\partial V(r(s),\lambda_s)}{\partial\lambda_s} .
\end{equation}

Finally, if we apply the transformation, Eq.~(\ref{trans}) to $x_c(s)$, then
\begin{align}
\tilde{x}_c(s)&=\frac 1 2 [x_+(t-s+i\hbar\beta/4)+x_-(t-s-i\hbar\beta/4)] \nonumber \\
&=r(t-s)+O(\hbar^2).
\end{align}
We therefore have in the classical limit,
\begin{equation}
\tilde{r}(s)=\tilde{x}_c(s)=r(t-s). \label{cl_trans1}
\end{equation} 
For $x_q$, we have
\begin{align}
\tilde{x}_q(s)&= x_+(t-s+i\hbar\beta/4)-x_-(t-s-i\hbar\beta/4) \nonumber \\
&=\hbar[\hat{x}(t-s) -\frac{i\beta}{2} \partial_s r(t-s) ]+O(\hbar^3).
\end{align}
On the other hand, we can write from Eq.~(\ref{xqxh})
$\tilde{x}_q(s)=\hbar[\tilde{\hat{x} }(s)+(i\beta/2)\partial_s \tilde{r}(s)]$.
Combining these two, we conclude that, in the classical limit,
\begin{equation}
\tilde{\hat{x}}(s)=\hat{x}(t-s)-i\beta\partial_s r(t-s). \label{cl_trans2}
\end{equation}
Equations (\ref{cl_trans1}) and (\ref{cl_trans2}) are  
exactly the set of  transformations used in the study of classical stochastic systems \cite{mallick,abc1,abc2,arenas} to obtain 
the classical fluctuation theorems.

\section{Summary and Discussion}

In summary, we have studied the path integral formulation of the quantum Brownian motion where the 
system particle is interacting with the environment consisting of a collection of bath harmonic oscillators.
We have developed the path integral on the deformed time contour following the procedure given in Ref.~\cite{abc},
and generalized to an open quantum system by deriving 
the expression for the influence functional which captures the effect of the environment on the system.
We then identified the field transformations for an open quantum system that leave the path integral action invariant 
in equilibrium. The transformations for the system variables take the same form as those 
found in closed systems \cite{sieb,abc}.
The open quantum system considered here is obtained by tracing out the bath degrees of freedom from the closed
system consisting of the system and the bath. In this paper, we have confirmed the expectation
that the same form of the transformations for the system variables does the job in the open system as 
those in closed systems by explicitly showing the invariance of the action and the influence functional when 
the system and the bath are initially at equilibrium. Since a general open quantum system can be regarded 
as a result of integrating out the bath degrees of freedom, we expect the same field transformations can
be used for general open systems.
This symmetry in equilibrium results in an identity for the two-time correlation function of the system
operators, which we show is just the fluctuation dissipation relation. 
When the system is driven by an external time-dependent protocol, the action is not invariant
under the transformations. In this non-equilibrium situation, using the change in the action under the transformations, 
we were able to find a version of quantum Jarzyski-type FT. In the process, 
we identified quantum work-like quantity defined 
on fluctuating quantum paths,
which reduces to the familiar expression for the work defined on a classical stochastic trajectory
in the classical limit.
Most importantly, we found that the action in the path integral formalism 
as well as the field transformations reduce in the classical limit to 
those used in the well-known MSRJD formalism for the classical generalized
Langevin equation. 

Exploring the symmetry and its breaking in classical stochastic systems has proven to be a quite useful tool
for understanding the equilibrium and non-equilibrium behavior of classical thermodynamic quantities \cite{mallick,abc1}. 
The present work provides a continuation of the previous efforts \cite{sieb,abc} to find the corresponding formalism
in the study of quantum stochastic thermodynamics. In this program, identifying the field
transformations is an important first step. The transformations must have the correspondence to 
the classical ones used in the MSRJD formalism and leave the action
and the influence functional invariant at equilibrium. As mentioned above, 
we have established the transformations for open systems, which can serve as
a starting point for the application of this formalism to more general non-equilibrium situations
occurring in an open quantum system where the symmetry is broken.
In this paper, we have only considered the case
where the system Hamiltonian is driven by the external agent, while the system plus bath are initially at equilibrium. 
The quantum work-like quantity we have found in this situation is quite similar to that found in Ref.~\cite{abc} for a
closed quantum system driven by an external protocol. In this case, the influence functional stays invariant under the
transformation. However, it is possible for the system to stay out of equilibrium even without the time-dependent driving
protocol. For example, it would be interesting to study the case where the initial state is given by the product state of 
the system and the bath or more generally by an arbitrary state. 
In this case, we expect the influence functional would {\it not} be invariant under the transformation.
In classical stochastic thermodynamics, one can identify the entropy production and heat from the bath by studying
the symmetry breaking term in the MSRJD functional \cite{abc1, arenas} in a similar situation. The explicit derivation 
of the influence functional and its symmetry property which we find in this work will be an essential 
starting point in 
investigating the entropy production and the heat in open quantum systems using the path integral approach. 
This is left for future study.

An important but difficult question is to find the physical observable corresponding to
the work-like quantity, Eq.~(\ref{workdef}) that arises in this paper. In closed systems, it
was associated with a time integral of an expectation value of an operator~\cite{abc}.
It is, however, hard to assign any simple physical meaning to this rather complicated combination of operators. 
It would be interesting to find the corresponding operator
for the open system and to investigate its physical meaning explicitly. This is also left for future study.

\onecolumngrid
\appendix
\section{Derivation of the influence functional $\Psi$}
\label{sec:appA}
In order to obtain Eq.~(\ref{psi}), we need to perform Gaussian path integrals over 
each bath variable $q_n(z)$ in Eq.~(\ref{normal_pi}) which are in general of the form,
\begin{equation}
 \int_{q_{n1}}^{q_{n2}}\mathcal{D}q_n(z) \;\exp[\frac i\hbar\int\limits_{\mathcal{C}} 
 dz\; \{ \frac{m_n}{2}(\frac{dq_n}{dz})^2-\frac 1 2
 m_n\omega_n^2q^2_n(z) +c_n x(z) q_n(z) \}],
\end{equation}
where $q_n(z)$ takes the values $q_{n1}$ and $q_{n2}$ at the start and end of a contour $\mathcal{C}$,
respectively. Let us first consider the case of $\mathcal{C}^\pm_2$ in Fig.~\ref{fig_one_point}, where
we have $z=s\pm i\hbar\beta/4$, $0\le s \le t$ 
and the integral over $z$ is just an integral over the real variable $s$. This type of 
path integral is well known \cite{feynman,kleinert}. For the contour $\mathcal{C}_2^+$ with the endpoints
$q_n(i\hbar\beta/4)=q^+_{n1}$ and $q_n(t+i\hbar\beta/4)=q^+_{n2}$, the result of the path integral,
denoted by $G_2^+$, is given by~\cite{kleinert}
\begin{equation}
 G_2^+=\left(\frac{m_n\omega_n}{2\pi i\hbar \sin (\omega_nt)}\right)^{1/2}\exp\Big[\frac i\hbar 
 \Phi(q^+_{n2},t;q^+_{n1},0\vert x_+)\Big],
\end{equation}
where, for a general time interval $(t_1,t_2)$,
\begin{align}
 &\Phi(q^+_{n2},t_2;q^+_{n1},t_1\vert x_+)=\frac{m_n \omega_n}
 {2\sin(\omega_n (t_2-t_1))}\Big[((q^{+}_{n1})^2+(q^{+}_{n2})^2)\cos(\omega_n (t_2-t_1)) 
 -2 q^+_{n1} q^+_{n2}\Big]  \label{phi}  \\
 &\quad\quad\quad ~~~ +\frac{c_n}{\sin(\omega_n (t_2-t_1))}\int_{t_1}^{t_2} ds\; \{q^+_{n2}\sin(\omega_n (s-t_1))
 +q^+_{n1}\sin(\omega_n(t_2-s))\} 
 x_+(s+i\hbar\beta/4) \nonumber \\
 &\quad\quad\quad ~~~ -\frac{c_n^2}{m_n\omega_n\sin(\omega_n (t_2-t_1))}\int_{t_1}^{t_2}\
 du\int_{t_1}^u ds\; \sin(\omega_n (s-t_1))\sin(\omega_n(t_2-u))
 x_+(u+i\hbar\beta/4)x_+(s+i\hbar\beta/4).
\nonumber
\end{align}
For the contour $\mathcal{C}^-_2$ which starts from $q_n(t-i\hbar\beta/4)=q^-_{n2}$ and ends at
$q_n(-i\hbar\beta/4)=q^-_{n1}$, the result now involves $x_-(s-i\hbar\beta/4)$ and can be similarly obtained 
from $\Phi(q^-_{n1},0;q^-_{n2},t\vert x_-)=-\Phi(q^-_{n2},t;q^-_{n1},0\vert x_-)$. We have the contribution 
from $\mathcal{C}^-_2$ as
\begin{equation}
 G_2^-=\left(\frac{m_n\omega_n}{-2\pi i\hbar \sin (\omega_nt)}\right)^{1/2}\exp\Big[-\frac i\hbar 
 \Phi(q^-_{n2},t;q^-_{n1},0\vert x_-)\Big],
\end{equation}
where the integral over the time now contains
$x_-(s-i\hbar\beta/4)$.

For the contours $\mathcal{C}^\pm_1$ and $\mathcal{C}^\pm_3$, the paths are along the imaginary axis. 
All these cases can be studied using the path which starts from $q_n(t_0+ib)=q''_{n}$ and ends at $q_n(t_0+ia)=q'_n$ with $a<b$.
For example, $t_0=0$ for $\mathcal{C}^\pm_1$ and $t_0=t$ for $\mathcal{C}^\pm_3$. 
The constants $a$ and $b$ can be read off 
from Fig.~\ref{fig_one_point}. The path integral is now 
with respect to the Euclidean action and can be calculated as \cite{kleinert}
\begin{align}
 &\int\mathcal{D}q_n \;\exp\Big[-\frac 1\hbar\int_a^b
 ds\; \big\{ \frac{m_n}{2}(\frac{dq_n}{ds})^2+\frac 1 2
 m_n\omega_n^2q^2_n(t_0+is) -c_n x(t_0+is) q_n(t_0+is) \big\}\Big] \nonumber \\
 =&\left(\frac{m_n\omega_n}{2\pi \hbar \sinh (\omega_n(b-a))}\right)^{1/2}
 \exp\Big[-\frac 1 \hbar
 \widetilde{\Phi}_{t_0}(q''_n,b;q'_n,a\vert x)\Big],
\end{align}
where
\begin{align}
 \widetilde{\Phi}_{t_0}(q''_n,b;q'_n,a\vert x)&=\frac{m_n \omega_n}{2\sinh(\omega_n (b-a))}\Big[(q''^{2}_{n}+q'^{2}_{n})\cosh(\omega_n (b-a)) 
 -2 q'_{n} q''_{n}\Big] \nonumber \\
 &-\frac{c_n}{\sinh(\omega_n (b-a))}\int_a^b ds\; \{q''_{n}\sinh(\omega_n (s-a))+q'_{n}\sinh(\omega_n(b-s))\} 
 x(t_0+is) \nonumber \\
 &-\frac{c_n^2}{m_n\omega_n\sinh(\omega_n (b-a))}\int_a^b du\int_a^u ds\; \sinh(\omega_n (s-a))\sinh(\omega_n(b-u))
 x(t_0+iu)x(t_0+is).
 \label{xi}
\end{align}
On the contours $\mathcal{C}^\pm_1$, we evaluate the path integrals over the bath variable with boundary conditions
$q_n(i\hbar\beta/2)=q_{ni}$, $q_n(i\hbar\beta/4)=q^+_{n1}$, $q(-i\hbar\beta/4)=q^-_{n1}$ and $q(-i\hbar\beta/2)=q_{ni}$.
The results are given by
\begin{equation}
G_1^+=\left(\frac{m_n\omega_n}{2\pi \hbar \sinh (\xi_n/4)}\right)^{1/2}
 \exp\Big[-\frac 1 \hbar
 \widetilde{\Phi}_0(q_{ni},\frac{\hbar\beta}2;q^+_{n1},\frac{\hbar\beta}4\vert x_+)\Big],
\end{equation}
and
\begin{equation}
G^-_1=\left(\frac{m_n\omega_n}{2\pi \hbar \sinh (\xi_n/4)}\right)^{1/2}
 \exp\Big[-\frac 1 \hbar
 \widetilde{\Phi}_0(q^-_{n1},-\frac{\hbar\beta}4;q_{ni},-\frac{\hbar\beta}2\vert x_-)\Big],
\end{equation}
where 
\begin{equation} 
\xi_n\equiv \beta\hbar\omega_n
\end{equation}
and $x_\pm$ appears inside the integral in Eq.~(\ref{xi}) as $x_\pm (\pm is)$.
Similarly, on $\mathcal{C}^\pm_3$, the boundary conditions are given by $q_n(t+i\hbar\beta/4)=q^+_{n2}$, $q_n(t)=q_{nf}$ and
$q_n(t-i\hbar\beta/4)=q^-_{n2}$ and the result of the path integral is
\begin{equation}
G^+_3=\left(\frac{m_n\omega_n}{2\pi \hbar \sinh (\xi_n/4)}\right)^{1/2}
 \exp\Big[-\frac 1 \hbar
 \widetilde{\Phi}_t(q^+_{n2},\frac{\hbar\beta}4;q_{nf},0\vert x_+)\Big],
\end{equation}
and
\begin{equation}
G^-_3=\left(\frac{m_n\omega_n}{2\pi \hbar \sinh (\xi_n/4)}\right)^{1/2}
 \exp\Big[-\frac 1 \hbar
 \widetilde{\Phi}_t(q_{nf},0;q^-_{n2},-\frac{\hbar\beta}4\vert x_-)\Big],
\end{equation}
where $x_\pm$ appears inside the integral in Eq.~(\ref{xi}) as $x_\pm (t\pm is)$.

The integral over the bath variable in Eq.~(\ref{normal_pi}) amounts to the evaluation of the following integral
\begin{equation}
I\equiv\prod_n\int_{-\infty}^\infty q^+_{n1}\int_{-\infty}^\infty q^+_{n2}
\int_{-\infty}^\infty q^-_{n1}\int_{-\infty}^\infty q^-_{n2}\int_{-\infty}^\infty q_{ni} \int_{-\infty}^\infty q_{nf}
\; G_1^+ G_2^+ G^+_3 G^-_3 G^-_2 G^-_1
\label{integral}
\end{equation}
All these integrals are Gaussian integral over real variables and can be done explicitly.
We first perform integrals over $q_{ni}$ and $q_{nf}$, respectively, as
\begin{align}
\int_{-\infty}^\infty dq_{ni}\; G_1^+ G^-_1 &= \left(\frac{m_n\omega_n}{2\pi \hbar \sinh (\xi_n/2)}\right)^{1/2} 
\exp\Big[ -\frac {m_n\omega_n} {2\hbar\sinh(\xi_n/2)} \Big\{ \cosh(\xi_n/2)((q^{+}_{n1})^2+(q^{-}_{n1})^2)
-2 q^+_{n1}q^-_{n1} \Big\} \nonumber \\
&\quad\quad 
+\frac{1}{\hbar\sinh(\xi_n/2)}\Big\{ q^+_{n1}(J_1+2\cosh(\xi_n/2)J^+_2)+q^-_{n1}(J_1+2\cosh(\xi_n/2)J^-_2)\Big\} \nonumber \\
&\quad\quad
+\frac{1}{m_n\omega_n\hbar\sinh(\xi_n/2)}\Big\{ \frac 1 2 J^2_1+2 \cosh(\xi_n/2) J_3\Big\}\Big],
\end{align}
and
\begin{align}
\int_{-\infty}^\infty dq_{nf}\; G^+_3 G^-_3 &= \left(\frac{m_n\omega_n}{2\pi \hbar \sinh (\xi_n/2)}\right)^{1/2} 
\exp\Big[ -\frac {m_n\omega_n} {2\hbar\sinh(\xi_n/2)} \Big\{ \cosh(\xi_n/2)((q^{+}_{n2})^2+(q^{-}_{n2})^2)
-2 q^+_{n2}q^-_{n2} \Big\} \nonumber \\
&\quad\quad 
+\frac{1}{\hbar\sinh(\xi_n/2)}\Big\{ q^+_{n2}(\tilde{J}_1+2\cosh(\xi_n/2)\tilde{J}^+_2)
+q^-_{n2}(\tilde{J}_1+2\cosh(\xi_n/2)\tilde{J}^-_2)\Big\} \nonumber \\
&\quad\quad
+\frac{1}{m_n\omega_n\hbar\sinh(\xi_n/2)}\Big\{ \frac 1 2 \tilde{J}^2_1+2 \cosh(\xi_n/2) \tilde{J}_3\Big\}\Big],
\end{align}
where
\begin{align}
&J_1=c_n\int_{\beta\hbar/4}^{\beta\hbar/2}ds\; \sinh(\omega_n s -\xi_n/4)(x_+(is)+x_-(-is)) \\
&J^\pm_2=c_n\int_{\beta\hbar/4}^{\beta\hbar/2}ds\; \sinh(\xi_n/2-\omega_n s)x_\pm (\pm is) \\
&J_3=c^2_n\int_{\beta\hbar/4}^{\beta\hbar/2}du \int_{\beta\hbar/4}^{u}ds\; \sinh(\xi_n/2-\omega_n u)
\sinh(\omega_n s -\xi_n/4)\{ x_+(iu)x_+(is)+x_-(-iu)x_-(-is)\},
\end{align}
and
\begin{align}
&\tilde{J}_1=c_n\int_0^{\beta\hbar/4}ds\; \sinh(\xi_n/4-\omega_n s)(x_+(t+is)+x_-(t-is)) \\
&\tilde{J}^\pm_2=c_n\int_0^{\beta\hbar/4}ds\; \sinh(\omega_n s)x_\pm (t\pm is) \\
&\tilde{J}_3=c^2_n\int_0^{\beta\hbar/4}du \int_0^{u}ds\; \sinh(\xi_n/4-\omega_n u)
\sinh(\omega_n s )\{ x_+(t+iu)x_+(t+is)+x_-(t-iu)x_-(t-is)\}.
\end{align}

The remaining integrals over $q^\pm_{n1}$ and $q^\pm_{n2}$ are again all Gaussian but not as straightforward
as the previous ones, since the variables are coupled.
We make a change of variables, with unit Jacobian, as
$Q_{n1}\equiv (1/2) (q^+_{n1}+q^-_{n1})$, $\bar{q}_{n1}\equiv q^+_{n1}-q^-_{n1}$, 
$Q_{n2}\equiv (1/2) (q^+_{n2}+q^-_{n2})$ and $\bar{q}_{n2}\equiv q^+_{n2}-q^-_{n2}$, then we find that the integrals over 
$\bar{q}_{n1}$ and $\bar{q}_{n2}$ can easily be done. We can then do the remaining integrals by decoupling the variables further
as $P_{n1}\equiv (1/\sqrt{2})(Q_{n1}+Q_{n2})$ and $P_{n2}\equiv (1/\sqrt{2})(Q_{n1}-Q_{n2})$.  The end results
for Eq.~(\ref{integral}) 
after a lengthy algebra 
can be summarised as
\begin{equation}
I=\prod_n\left(\frac{1}{2\sinh(\xi_n/2)}\right)\exp[-\frac 1\hbar \sum_{a=1}^6\Psi^{(a)}[x_+,x_-]],
\label{i-result}
\end{equation}
where we have separated the contributions into six parts in such a way that $\Psi^{(1)}$ involves 
the integrals over the real time axis only and 
$\Psi^{(2)}, \Psi^{(3)}$ and $\Psi^{(4)}$ contain those over the imaginary time axis.
The remaining $\Psi^{(5)}$ and $\Psi^{(6)}$ are given as mixed time integrals over the real and imaginary time axes.
We obtain
\begin{align}
 \Psi^{(1)}[x_+,x_-]=\sum_n\frac{c^2_n}{2m_n\omega_n\sinh(\xi_n/2)}&\Big[ \int_0^tdu\int_0^u ds\; 
 \{ x_+(u+i\hbar\beta/4)\cos(\omega_n(u-s)+i\xi_n/2) x_+(s+i\hbar\beta/4) \nonumber \\
 &\quad\quad\quad\quad  +x_-(u-i\hbar\beta/4)\cos(\omega_n(u-s)-i\xi_n/2) x_-(s-i\hbar\beta/4)\} \nonumber \\
 &-\int_0^tdu\int_0^t ds\; x_+(u+i\hbar\beta/4)\cos(\omega_n(u-s)) x_-(s-i\hbar\beta/4) \Big],
\end{align}
\begin{align}
 \Psi^{(2)}[x_+,x_-]=-\sum_n&\frac{c^2_n}{2m_n\omega_n\sinh(\xi_n/2)}\Big[
 \int_{\beta\hbar/4}^{\beta\hbar/2}du\int_{\beta\hbar/4}^{\beta\hbar/2} ds\; \cosh(\xi_n/2-\omega_n(u+s))
 x_-(-iu)x_+(is) \\
 &+ \int_{\beta\hbar/4}^{\beta\hbar/2}
 du\int_{\beta\hbar/4}^u ds\; \cosh(\xi_n/2-\omega_n(u-s)) \{x_+(iu)x_+(is)+x_-(-iu)x_-(-is)\} \Big], \nonumber 
 \end{align}
 \begin{align}
 \Psi^{(3)}[x_+,x_-]=-\sum_n&\frac{c^2_n}{2m_n\omega_n\sinh(\xi_n/2)}\Big[
 \int_{0}^{\beta\hbar/4}du\int_{0}^{\beta\hbar/4} ds\; \cosh(\xi_n/2-\omega_n(u+s))
 x_-(t-iu)x_+(t+is) \\
 &+\int_{0}^{\beta\hbar/4}
 du\int_{0}^u ds\; \cosh(\xi_n/2-\omega_n(u-s)) \{x_+(t+iu)x_+(t+is)+x_-(t-iu)x_-(t-is)\}  \Big], \nonumber 
 \end{align}
 \begin{align}
 \Psi^{(4)}[x_+,x_-]=-\sum_n\frac{c^2_n}{2m_n\omega_n\sinh(\xi_n/2)}\Big[&
 \int_{\beta\hbar/4}^{\beta\hbar/2}du\int_0^{\beta\hbar/4} ds\; \{\cosh(\xi_n/2-\omega_n(u-s)-i\omega_n t)
 x_+(iu)x_+(t+is) \nonumber \\
 &+\cosh(\xi_n/2-\omega_n(u+s)-i\omega_n t)x_+(iu)x_-(t-is) \nonumber \\
 &+\cosh(\xi_n/2-\omega_n(u+s)+i\omega_n t)x_-(-iu)x_+(t+is) \nonumber \\
 &+\cosh(\xi_n/2-\omega_n(u-s)+i\omega_n t)x_-(-iu)x_-(t-is) \} \Big],
\end{align}
\begin{align}
 \Psi^{(5)}[x_+,x_-]=-i\sum_n\frac{c^2_n}{2m_n\omega_n\sinh(\xi_n/2)} \Big[ &
 \int_{\beta\hbar/4}^{\beta\hbar/2} du\int_0^t ds\; 
 \{ \cosh(3\xi_n/4-i\omega_n(s-iu)) x_+(iu)x_+(s+i\hbar\beta/4) \nonumber \\
 & -\cosh(\xi_n/4-i\omega_n(s-iu))x_+(iu) x_-(s-i\hbar\beta/4) \nonumber \\
 &+\cosh(\xi_n/4+i\omega_n(s+iu)) x_-(-iu)x_+(s+i\hbar\beta/4) \nonumber \\
&-\cosh(3\xi_n/4+i\omega_n(s+iu)) x_-(-iu)x_-(s-i\hbar\beta/4) \} \Big],
 \end{align}
and
\begin{align}
 \Psi^{(6)}[x_+,x_-]=-i\sum_n\frac{c^2_n}{2m_n\omega_n\sinh(\xi_n/2)} \Big[ &
 \int_{0}^{\beta\hbar/4} du\int_0^t ds\; 
 \{ \cosh(\xi_n/4-i\omega_n(t-s+iu)) x_+(t+iu)x_+(s+i\hbar\beta/4) \nonumber \\
 & -\cosh(\xi_n/4+i\omega_n(t-s+iu))x_+(t+iu) x_-(s-i\hbar\beta/4) \nonumber \\
 &+\cosh(\xi_n/4-i\omega_n(t-s-iu)) x_-(t-iu)x_+(s+i\hbar\beta/4) \nonumber \\
&-\cosh(\xi_n/4+i\omega_n(t-s-iu)) x_-(t-iu)x_-(s-i\hbar\beta/4) \} \Big].
 \end{align}
We note that the prefactor in Eq.~(\ref{i-result}) is just $Z_B$ in Eq.~(\ref{zb}), which combined with $Z_\beta(0)$ produces
$1/Z(0)$ in Eq.~({\ref{S_ave1}). We can easily see that the above expressions for $\Psi^{(a)}$ can be rewritten 
in a more compact way as double contour integrals in Eq.~(\ref{psi}). We also note that the terms that depend on $\mu$
in Eq.~(\ref{psi}) follow from the interaction Lagrangian Eq.~(\ref{LI}) not from the Gaussian integration of the bath variables.

\section{Invariance of $\Psi$ under the transformation Eq.~(\ref{trans})}
\label{sec:appB}

We show that $\Psi[\tilde{x}_+,\tilde{x}_-]=\Psi[x_+,x_-]$ where $\tilde{x}_\pm$
is given by Eq.~(\ref{trans}). We look at how each $\Psi^{(a)}$, $a=1,2,\cdots,6$, 
defined in Appendix \ref{sec:appA} changes 
under this transformation.
We first express $\Psi^{(1)}[\tilde{x}_+,\tilde{x}_-]$ in terms of $x_\pm$
using Eq.~(\ref{trans}) and then it is a simple exercise of change of integration variables, 
$u\to t-u$ and $s\to t-s$ to show that $\Psi^{(1)}[\tilde{x}_+,\tilde{x}_-]=\Psi^{(1)}[x_+,x_-]$. 
In this and the following calculations, we use $\int_0^t du\int_u^t ds=\int_0^tds\int_0^s du$ and similar 
expressions for a double integral. For $\Psi^{(2)}[\tilde{x}_+,\tilde{x}_-]$, we change the integration variables as 
$u\to \beta\hbar/2-u$ and $s\to \beta\hbar/2-s$. We then find 
$\Psi^{(2)}[\tilde{x}_+,\tilde{x}_-]=\Psi^{(3)}[x_+,x_-]$. Similarly, we have 
$\Psi^{(3)}[\tilde{x}_+,\tilde{x}_-]=\Psi^{(2)}[x_+,x_-]$. By the same variable change, we have 
$\Psi^{(4)}[\tilde{x}_+,\tilde{x}_-]=\Psi^{(4)}[x_+,x_-]$. For $\Psi^{(5)}[\tilde{x}_+,\tilde{x}_-]$
and $\Psi^{(6)}[\tilde{x}_+,\tilde{x}_-]$, we use the change of variables, $u\to \beta\hbar/2-u$
and $s\to t-s$. We can easily check that 
$\Psi^{(5)}[\tilde{x}_+,\tilde{x}_-]=\Psi^{(6)}[x_+,x_-]$ and
$\Psi^{(6)}[\tilde{x}_+,\tilde{x}_-]=\Psi^{(5)}[x_+,x_-]$. The $\mu$-dependent term is also invariant 
for the same reason as explained in the main text. Therefore, we conclude that
$\Psi[\tilde{x}_+,\tilde{x}_-]=\Psi[x_+,x_-]$.

\section{Integration over the bath variables in the calculation of correlation functions}
\label{sec:appC}

Here we focus on the calculation of the matrix elements arising in the calculation of the correlation functions
given in Eq.~(\ref{corrB}) as
\begin{equation}
 \int d\bar{\bm{q}}\;\langle x'_2,\bm{q}'_2\vert e^{\beta H_{\rm tot}/4} 
\vert \bar{x}',\bar{\bm{q}}\rangle 
 \langle \bar{x},\bar{\bm{q}}\vert e^{-\beta H_{\rm tot}/4} \vert
x_2,\bm{q}_2\rangle .
\label{matrix-el}
\end{equation}
We will show below that when we perform the path integral over the bath variables, we obtain the delta function $\delta(\bm{q}_2-\bm{q}'_2)$.
We first express these matrix elements using the path integral representation. The paths are along the imaginary time axis
with the real part $t_2$.
The path integral over each bath variable $q_n(z)$ is Gaussian and can be evaluated explicitly as we have done in Appendix \ref{sec:appA}.
When integrated over $q_n$, the second matrix element in Eq.~(\ref{matrix-el}) gives a factor of 
\begin{equation}
\left(\frac{m_n\omega_n}{2\pi \hbar \sinh (\xi_n/4)}\right)^{1/2}
 \exp\Big[-\frac 1 \hbar
 \widetilde{\Phi}_{t_2}(q_{n2},\frac{\beta\hbar}4;\bar{q}_n,0\vert x)\Big],
 \label{f1}
\end{equation}
while the first one gives
\begin{equation}
\left(\frac{m_n\omega_n}{-2\pi \hbar \sinh (\xi_n/4)}\right)^{1/2}
 \exp\Big[-\frac 1 \hbar
 \widetilde{\Phi}_{t_2}(\bar{q}_n,0;q'_{n2},\frac{\beta\hbar}4\vert x)\Big].
 \label{f2}
\end{equation}
where $\widetilde{\Phi}$ is given in Eq.~(\ref{xi}) and $\xi_n=\beta\hbar\omega_n$.
We note that 
$ \widetilde{\Phi}_{t_2}(\bar{q}_n,0;q'_{n2},\beta\hbar/4\vert x)=
-  \widetilde{\Phi}_{t_2}(q'_{n2},\beta\hbar/4;\bar{q}_n,0\vert x)$.

If we multiply these two factors and try to integrate over $\bar{q}_n$, we find that the integral looks ill-defined. 
We can, however, regularize this integral by putting slightly different $\beta'$ in Eq.~(\ref{f2}) from $\beta$ in Eq.~(\ref{f1}) 
and taking the $\beta'\to\beta$ limit in the end. Therefore, apart from the prefactors in Eqs.~(\ref{f1}) and (\ref{f2}), 
we have to evaluate the integral,
\begin{align}
\int_{-\infty}^\infty d\bar{q}_{n}\; \exp\Big[ &-\frac{m_n\omega_n}{2\hbar}\{\coth(\xi_n/4)-\coth(\xi'_n/4)\}\bar{q}^2_n 
+\frac{m_n\omega_n}{\hbar}\{\frac{q_{n2}}{\sinh(\xi_n/4)}-\frac{q'_{n2}}{\sinh(\xi'_n/4)}\}\bar{q}_n \nonumber \\
&+\frac 1\hbar (I(\xi_n)-I(\xi'_n))\bar{q}_n -\frac{m_n\omega_n}{2\hbar}\{\coth(\xi_n/4)q^2_{n2}-\coth(\xi'_n/4)q'^2_{n2}\} \nonumber \\
&+\frac 1 \hbar (J(\xi_n)q_{n2}-J(\xi'_n)q'_{n2}) +O(\xi_n-\xi'_n) \Big],
\end{align}
where $\xi'_n=\beta'\hbar\omega_n$ and 
\begin{align}
&I(\xi_n) = \frac{c_n}{\sinh(\xi_n/4)}\int_0^{\beta\hbar/4} ds\; \sinh(\frac{\xi_n}4-s)x(t_2+is) \\
&J(\xi_n)= \frac{c_n}{\sinh(\xi_n/4)}\int_0^{\beta\hbar/4} ds\; \sinh(s)x(t_2+is)
\end{align}
with the corresponding primed expressions.
We note that we have neglected the terms coming from the
double integrals in Eq.~(\ref{xi}) which vanish in the limit $\xi'_n\to \xi_n$.

We now perform the Gaussian integral over $\bar{q}_n$. The prefactor coming from this integral combined with two prefactors
in Eqs.~(\ref{f1}) and (\ref{f2}) gives $(m_n\omega_n/(2\pi\hbar \sinh((\xi_n-\xi'_n)/4)))^{1/2}$.  We also need a relation
\begin{equation}
I(\xi_n)-I(\xi'_n)=\frac{\xi_n-\xi'_n}{4}\frac {J(\xi_n)}{\sinh(\xi_n/4)} +O((\xi_n-\xi'_n)^2).
\end{equation}
Combining all these terms, we find that the result of the path integral over the bath variable gives us
\begin{equation}
\left(\frac{m_n\omega_n}{2\pi\hbar \sinh((\xi_n-\xi'_n)/4)}\right)^{1/2}\exp\Big[-\frac{m_n\omega_n}{2\hbar}
\frac{1}{\sinh((\xi_n-\xi'_n)/4)}\Big\{ (q_{n2}-q'_{n2})^2 +O((\xi_n-\xi'_n)^2)\Big\} \Big].
\end{equation}
In the limit $\xi'_n\to \xi_n$, the above expression reduces to $\delta(q_{n2}-q'_{n2})$.

\begin{acknowledgments}
The author was supported by
Basic Science Research Program through the National Research Foundation 
of Korea (NRF) funded by the Ministry of Education (2017R1D1A09000527).
\end{acknowledgments}
\twocolumngrid

\end{document}